\documentstyle[sprocl]{article}

\input{psfig}

\def\Journal#1#2#3#4{{#1} {\bf #2}, #3 (#4)}

\def\ANN{\em Ann. Phys.}
\def\ANP{\em Adv. Nucl. Phys.}
\def\ARNPS{\rm Ann. Rev. Nucl. Part. Sci.}
\def\EPJA{{\em Eur. Phys. J.} A}
\def\EPJC{{\em Eur. Phys. J.} C}
\def\JPG{{\em J. Phys.} G}
\def\NPA{{\em Nucl. Phys.} A}
\def\NPB{{\em Nucl. Phys.} B}
\def\PL{\em Phys. Lett.}
\def\PLB{{\em Phys. Lett.} B}
\def\PR{\em Phys. Rev.}
\def\PRL{\em Phys. Rev. Lett.}
\def\PRC{{\em Phys. Rev.} C}
\def\PRD{{\em Phys. Rev.} D}
\def\PRP{\em Phys. Rep.}
\def\PPNP{\em Prog. Part. Nucl. Phys.}
\def\RMP{\em Rev. Mod. Phys.}
\def\SJNP{\em Sov. J. Nucl. Phys.}
\def\JETP{\em Sov. Phys. JETP}
\def\ZPA{{\em Z. Phys.} A}
\def\ZPC{{\em Z. Phys.} C}

\begin{document}

\hspace*{5cm} ADP-00-31/T414,\ \ JLAB-THY-00-18 \\

\title{QCD and the Structure of the Nucleon in Electron Scattering%
	\footnote{Lectures presented at the 1999 Hampton University
		Graduate Studies (HUGS) summer school,
		Jefferson Lab.}}

\author{W. MELNITCHOUK}

\address{Special Research Centre for the
	Subatomic Structure of Matter,  \\
	University of Adelaide, Adelaide 5005, Australia, and\\
	Jefferson Lab, 12000 Jefferson Avenue,
	Newport News, VA 23606}

\maketitle

\abstracts{The internal structure of the nucleon is discussed within the
	context of QCD.
	Recent progress in understanding the distribution of flavor and
	spin in the nucleon is reviewed, and prospects for extending our
	knowledge of nucleon structure in electron scattering experiments
	at modern facilities such as Jefferson Lab are outlined.}

\tableofcontents{}
\newpage

\section{Introduction}

The internal structure of the nucleon is the most fundamental problem
of strong interaction physics.
Understanding this structure in terms of the elementary quark and gluon
degrees of freedom of the underlying theory, quantum chromodynamics
(QCD), remains the greatest unsolved problem of the Standard Model of
nuclear and particle physics.

Historically, the basic strong interaction which we have sought to
explain has been that between protons and neutrons in the atomic nucleus.
The original idea of massive particle exchange of Yukawa~\cite{YUKAWA}
has been a guiding principle according to which later theories have been
developed.
It was pointed out by Wick~\cite{WICK} that this idea was consistent with
the Heisenberg Uncertainty principle, whereby the interaction range of
the nuclear force is inversely proportional to the mass of the exchanged 
meson.
Over the years a phenomenological description of the forces acting
between nucleons has been developed within a meson-exchange picture.

Following the experimental discovery of the pion in 1947, the 1950s
and 1960s saw an explosion of newly discovered mesons and baryons,
as particle accelerators pushed to higher energies.
To bring some sense of order to the profusion of new particles, Gell-Mann
and Zweig introduced the idea of quarks~\cite{QUARKS}, which enabled much
of the hadronic spectrum to be organized in terms of just a few elementary
constituents.
Soon after, however, it was realized that a serious problem existed with
the simple quark classifications, namely the $\Delta^{++}$ isobar.
The quark model wave function for the $\Delta^{++}$ was predicted to
be totally symmetric, however the $\Delta^{++}$ obeyed Fermi-Dirac
statistics.
A solution to this problem was found by assigning extra internal
{\em color}~\cite{COLOR} quantum numbers to the quarks, in which
baryons would have in addition an antisymmetric color wave function.

The discovery of scaling in deep-inelastic electron--nucleon scattering
in the late 1960s at SLAC~\cite{EARLYSLAC} confirmed that the nucleon
contained point-like constituents, which were soon identified with the
quarks of the quark model.
Imposing local gauge invariance on the color fields, and introducing
vector gluon exchange to mediate the inter-quark interaction, led
naturally to the development of QCD as the fundamental theory of strong
interactions~\cite{QCD}.

Because QCD is an asymptotically free theory --- the effective strong
coupling constant decreases at short distances --- processes involving
large momentum transfers can be calculated reliably within perturbation
theory.
Yet despite the successes of perturbative QCD, we are still unable to
extract from QCD sufficient details regarding its long-distance
properties.
This is because in the infra-red region the strong coupling constant
grows and perturbation theory breaks down, and the available
non-perturbative tools are not yet sufficiently developed to allow
quantitative predictions.

In a sense it is ironic that the theory which arose out of the desire
to understand nuclear forces is able to explain backgrounds in hadronic
jets produced in high energy collisions, yet is unable to describe the
properties of the ground state of the theory.
Although one can argue that QCD in principle explains all hadronic
and nuclear phenomena, without understanding the consequences of QCD
for hadron phenomenology one may as well argue that the entire physics
of atoms and molecules can in principle be explained from 
QED~\cite{MOLECULE}.
Understanding how the transition from the quarks and gluons of QCD to
the physical mesons and baryons takes place remains the holy grail of
modern nuclear physics.

In the next Section some basic elements of QCD relevant for later
application to nucleon structure are reviewed.
This is followed in Section~3 by the basic definitions and kinematics
of electron--nucleon scattering, including elastic, deep-inelastic
and semi-inclusive scattering.
In Section~4 we focus more closely on the flavor and spin content
of the nucleon, and outline some recent highlights in the study of
valence and sea quark distributions.
Finally, some concluding remarks are made in Section~5.

\vspace*{0.5cm}

\section{Elements of QCD}

Quantum Chromodynamics is a non-Abelian gauge field theory based on the
gauge group SU(3)$_{\rm color}$, and defined in terms of the Lagrange
density~$^{6,8-10}$:
\begin{eqnarray}
\label{LQCD}
{\cal L}_{\rm QCD}
&=& {\cal L}_{\rm inv}\
 +\ {\cal L}_{\rm gauge}\
 +\ {\cal L}_{\rm ghost}\ ,
\end{eqnarray}
where ${\cal L}_{\rm inv}$ is the classical Lagrangian, invariant under
local gauge transformations of the SU(3)$_{\rm color}$ group:
\begin{eqnarray}
\label{Linv}
{\cal L}_{\rm inv}
&=& \overline\psi_{i, f} (i \gamma^\mu D_{\mu} - m_f)_{ij}
    \psi_{j, f}\
 -\ {1\over 4} F_{\mu\nu}^a F^{\mu\nu\ a}\ .
\end{eqnarray}
The quark fields $\psi_{i, f}$ (for a particular quark flavor
$f = u, d, s, \cdots$\ , with mass $m_f$) are labeled by color indices
$i,j=1,2,3$.
The covariant derivative is $D_\mu = \partial_\mu - i g T^a A_{\mu}^a$,
where $g$ is the QCD coupling constant and $T^a$ are the generators of
the SU(3) group, with $a=1, \cdots, 8$.
In terms of the gluon field $A_\mu^a$ the gluon field strength tensor is
$F_{\mu\nu}^a = \partial_\mu A_\nu^a - \partial_\nu A_\mu^a
		+ g f^{abc} A_\mu^b A_\nu^c$,
with $f^{abc}$ the SU(3) structure constants.
The major differences between QCD and quantum electrodynamics is the
appearance, due to the non-Abelian structure of the theory, of gluon
self-couplings in the $F\cdot F$ term.
This gives rise to 3- and 4-point gluon interactions which make the
theory highly non-linear, but also leads to the property of asymptotic
freedom (Section~3.2).

Under local gauge rotations the quark fields transform (dropping color
and flavor indices) according to:
\begin{eqnarray}
\psi(x) &\rightarrow& \psi'(x)\ =\ U(x)\ \psi(x)\ ,
\end{eqnarray}
where $U$ is an SU(3) unitary matrix:
\begin{eqnarray}
U(x) &=& \exp\left( i \theta^a(x) T^a \right)\ ,
\end{eqnarray}
with $\theta^a(x)$ real.
The gluon fields transform according to:
\begin{eqnarray}
A_\mu(x) &\rightarrow& A'_\mu(x)\
 =\ U(x) A_\mu(x) U^{-1}(x)
 + { i \over g } \left( \partial_\mu U(x) \right) U^{-1}(x)\ .
\end{eqnarray}
One can easily show that ${\cal L}_{\rm inv}$ is then invariant under
these transformations.

The gauge invariance of the classical Lagrangian introduces some
difficulties when quantizing the gauge theory.
This problem is avoided with the introduction of an addition term,
${\cal L}_{\rm gauge}$, which fixes a specific gauge.
In the Lorentz (covariant) gauge, one has
${\cal L}_{\rm gauge} = -(1/2\alpha) (\partial^\mu A^a_\mu)^2$,
where $\alpha$ is an arbitrary gauge parameter.
Of course observables cannot depend on the choice of $\alpha$,
and some common choices are $\alpha=1$ (Feynman gauge) and
$\alpha\rightarrow 0$ (Landau gauge).
Other, non-covariant, gauge choices are the Coulomb gauge
($\partial_i A^a_i=0$), the axial gauge ($A^a_3=0$),
and the temporal gauge ($A^a_0=0$).

The Faddeev-Popov ghost density,
${\cal L}_{\rm ghost}
= (\partial_\mu \bar \chi^a)
  (\delta^{ab} \partial_\mu - g f^{abc} A_\mu^c) \chi^c$,
where $\chi^a$ and $\bar \chi^a$ are scalar {\em anti}-commuting
ghost fields, ensures that the gauge fixing does not spoil the
unitarity of the $S$-matrix.
Further discussion about the gauge fixing problem can be found
in Refs.~\cite{MUTA,HANDBOOK}.

In renormalizable field theories such as QCD the strength of the
interaction depends on the energy scale.
A property almost unique to QCD is that the renormalized coupling constant
decreases with energy --- known as asymptotic freedom.
The running of the QCD coupling with energy allows one to compute cross
sections for any quark--gluon process using a perturbative expansion
if the coupling is small.
Writing the full QCD Lagrangian as\
${\cal L}_{\rm QCD} = {\cal L}_0 + {\cal L}_{\rm I}$, where
\newpage
\begin{eqnarray}
\label{L0}
{\cal L}_0 &=&
\overline\psi (i \gamma^\mu \partial_\mu - m) \psi
- {1 \over 4}
  (\partial_\mu A^a_\nu - \partial_\nu A^a_\mu)
  (\partial^\mu A^{a\ \nu} - \partial^\nu A^{a\ \mu})	\nonumber\\
& &
- {1 \over 2\alpha} (\partial^\mu A^a_\mu)^2
+ (\partial^\mu \bar\chi^a) (\partial_\mu \chi^a)\
\end{eqnarray}
is the free Lagrangian, and
\begin{eqnarray}
\label{Lint}
{\cal L}_{\rm I}
&=& g\ \overline\psi \gamma^\mu T^a \psi A^a_\mu
- {g \over 2} f^{abc}
  (\partial_\mu A^a_\nu - \partial_\nu A^a_\mu) A^{b \mu} A^{c \nu}
	\nonumber\\
& &
- {g^2 \over 4} f^{abe} f^{cde} A^a_\mu A^b_\nu A^{c \mu} A^{b \nu}
- g f^{abc} (\partial^\mu \bar\chi^a) \chi^b A^c_\mu
\end{eqnarray}
the interaction part, one can derive from ${\cal L}_{\rm I}$
a complete set of Feynman rules for computing any scattering
amplitude involving quarks and gluons~$^{8-11}$.

Perturbative QCD has been enormously successful in calculating hard
processes in high energy lepton-lepton, lepton-hadron and hadron-hadron
scattering.
However, even at high energies one can never avoid the fact that the
physical states from which the quarks and gluons emerge to undergo the
hard scattering are hadrons, so that one always encounters soft scales
in any strongly interacting system.
While operator product expansions usually allow one to factorize the
short and long distance dynamics, understanding the complete physical
process necessarily requires going beyond perturbation theory.

Over the years the problem of non-perturbative QCD has been tackled
on several fronts.
The most direct way is to solve the QCD equations of motion numerically
on a discretized space-time lattice~\cite{LATTICE}.
Recent advances in lattice gauge field theory and computing power has
made quantitative comparison of full lattice QCD calculations with
observables within reach.

Alternative methods of tackling non-perturbative QCD involve the building
of soluble, low-energy QCD-inspired models, which incorporate some, but
not all, of the elements of QCD.
Phenomenological input is then used to constrain the model parameters,
and identify circumstances where various approximations may be
appropriate.
These approaches often exploit specific symmetries of QCD, which for
some observables may bring out the essential aspects of the physics
independent of the approximations used elsewhere.
A good example of this, which has had extensive applications in low
energy physics, is chiral symmetry.

Consider the classical quark Lagrange density in Eq.(\ref{Linv}) in the
limit where the mass of the quarks is zero:
\newpage
\begin{eqnarray}
\label{Lchiral}
{\cal L}^q_{\rm inv}
&=& \overline\psi i \gamma^\mu D_\mu \psi\
 =\ \overline\psi_L i \gamma^\mu D_\mu \psi_L\
 +\ \overline\psi_R i \gamma^\mu D_\mu \psi_R\ ,
\end{eqnarray}
where 
$\psi_{L,R} = (1/2) (1 \pm \gamma_5) \psi$ are left- and right-handed
projections of the Dirac fields.
Under independent global left- and right-handed rotations
${\cal L}^q_{\rm inv}$ remains unchanged.
For $N_f$ massless quarks, the classical QCD Lagrangian is then said to
have a chiral SU$(N_f)_L \otimes$~SU$(N_f)_R$ symmetry.
Of course non-zero quark masses break this symmetry explicitly by mixing
left- and right-handed quark fields, however, for the $u$ and $d$ quarks,
and to some extent the $s$, the masses are small enough for the chiral
symmetry to be approximately valid.

If chiral symmetry were exact, a natural consequence would be parity
doubling.
The nucleon would have a negative parity partner with the same mass,
and the pseudoscalar pion would have the same mass as the scalar meson.
In nature, the lightest negative parity spin-1/2 baryon is the $S_{11}$
resonance, which is several hundred MeV heavier than the nucleon.
Moreover, the pion has an exceptionally small mass, while the lightest
candidate for a scalar meson is several times heavier than the pion,
so that chiral symmetry in nature is clearly broken.

The way to reconcile a symmetry which is respected by the Lagrangian
but broken by the physical ground state is if the symmetry is broken
spontaneously.
According to Goldstone's theorem, a consequence of a spontaneously broken
chiral symmetry is the appearance of massless pseudoscalar bosons.
For $N_f=2$ (namely, for $u$ and $d$ flavors), these correspond to the
pseudoscalar pions; for $N_f=3$ (counting the strange quark as light),
these include in addition the kaons and the $\eta$ meson.
The physical mesons are of course not massless, but on the scale of
typical hadronic masses ($\sim 1$~GeV), they can be considered light
--- their masses arising from the small but non-zero quark masses.
A perturbative expansion in terms of the small pseudoscalar boson masses
can be developed~\cite{CHIPT}, and applied systematically to describe
hadron interactions at very low energies.

As will be demonstrated in Sections~4.2 and 4.3 the chiral properties
of QCD are in fact critical to understanding many aspects of nucleon
structure, from low energy form factors to deep-inelastic structure
functions.
Before proceedings with further discussion about QCD and nucleon
structure, however, in the next Section we first define the observables
through which one can study the internal structure of the nucleon in
electron scattering.

\newpage
\section{Electron--Nucleon Scattering}

Because the electromagnetic interaction of leptons is perhaps the best
understood part of the Standard Model, the cleanest way to probe the
internal structure of hadrons is through lepton scattering.
This applies to both charged leptons and neutrinos, although to be
concrete we shall consider the scattering of electrons.

In the one-photon approximation, the scattering of an electron with
four-momentum $l$ from a nucleon with momentum $p$ is illustrated in
Fig.~1, where the outgoing electron momentum is $l'$
and the hadronic final state is denoted by $X$:\ $e N \rightarrow e' X$.
The energies of the incident and scattered electrons are $E$ and $E'$,
and the electron scattering angle is $\theta$.
The energy transfer to the nucleon in the target rest frame is
$\nu = E - E'$, and the four-momentum transfer is
$q^2 = (l-l')^2 \simeq -4 E E' \sin^2\theta$ for $m_e \ll E,E'$.

In the nucleon rest frame the differential cross section is given by:
\begin{eqnarray}
\label{sigmadef}
{d^2 \sigma \over d\Omega dE'}
&=& { \alpha^2_{em} \over 2 M Q^4 } { E' \over E }\
    L^{\mu\nu}\ W_{\mu\nu}\ ,
\end{eqnarray}
where $\alpha_{em}$ is the electromagnetic fine structure constant,
$M$ is the nucleon mass, and $Q^2 \equiv -q^2$.
The normalization of states in Eq.(\ref{sigmadef}) is such that
$\langle \vec p | \vec p\ ' \rangle
 = (2\pi)^3\ 2 p_0\ \delta(\vec p - \vec p\ ')$.
The lepton tensor $L^{\mu\nu}$ is given by:
\begin{eqnarray}
L^{\mu\nu}
&=& 2\ l^{\mu} l'^{\nu} + 2\ l'^{\mu} l^{\nu} + g^{\mu\nu} q^2\
 \mp\ 2i \epsilon^{\mu\nu\lambda\rho} l_\lambda l'_\rho\ ,
\end{eqnarray}
corresponding to an electron with helicity $\pm 1/2$.
Note that for unpolarized scattering only the part of $L^{\mu\nu}$
which is symmetric under the interchange $\mu \leftrightarrow \nu$
is relevant, while for polarized only the antisymmetric part enters.

The hadronic tensor,
\begin{eqnarray}
\label{wmunudef}
W_{\mu\nu}
&=& {1\over 2} \sum_X (2\pi)^3 \delta^4(p+q-p_X)\
             \langle N | J_{\mu}(0) | X \rangle
             \langle X | J_{\nu}(0) | N \rangle\ ,
\end{eqnarray}
contains all of the information on the structure of the target.
For inclusive scattering one sums over all final states $X$,
while for exclusive scattering $X$ denotes a specific hadron.

The most general form for the hadronic tensor consistent with
Lorentz and gauge invariance, as well as invariance under time
reversal and parity, is:
\begin{eqnarray}
{1 \over 2M} W_{\mu\nu}
&=& \left( -g_{\mu\nu} + { q_\mu q_\nu \over q^2 } \right) W_1
 +\ \left( p_\mu - {p \cdot q \over q^2} q_\mu \right)
    \left( p_\nu - {p \cdot q \over q^2} q_\nu \right) W_2   \nonumber\\
&+& i \epsilon_{\mu\nu\lambda\rho} {q^\lambda \over M}
	\left( s^\rho M^2 G_1
	     + (p\cdot q\ s^\rho - s\cdot q\ p^\rho) G_2
	\right)\ ,
\end{eqnarray}
where $s^\rho$ is the nucleon spin vector, defined by
$s^\rho(\lambda) = (2\lambda/M) (|\vec p\ |; p_0 \hat p)$
for nucleon helicity $\lambda$, so that $s^2=-1$, $s \cdot p=0$.
The structure functions $W_1$, $W_2$, $G_1$ and $G_2$ are in general
functions of two variables.
Usually one chooses these to be $Q^2$, and the Bjorken $x$ variable,
defined as $x = Q^2/2 p\cdot q$.

\begin{figure}
\label{fig:dis}
\centering{\ \psfig{figure=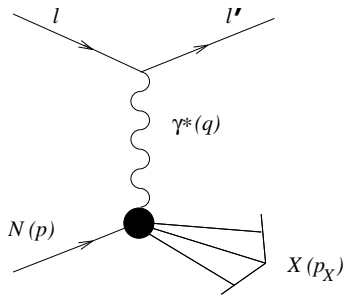,height=4cm}}
\caption{Electron--nucleon scattering in the one-photon exchange
	approximation: $X=N$ for elastic scattering,
	while $X$ is a sum over hadrons for inclusive inelastic
	scattering.}
\end{figure}

Since $W_1$ and $W_2$ are coefficients of Lorentz tensors symmetric
in $\mu\nu$, they can be measured in unpolarized electron--nucleon
scattering.
The unpolarized differential cross section can be written in the target
rest frame as~\cite{HEY,LP}:
\begin{eqnarray}
\label{sigdis}
{ d^2\sigma^{\uparrow\Uparrow+\downarrow\Uparrow} \over d\Omega dE' }
&=& { 8 \alpha^2 E'^2 \over Q^4 }
    \left[ 2 W_1(x,Q^2) \sin^2(\theta/2) + W_2(x,Q^2) \right]\ ,
\end{eqnarray}
where $\uparrow\Uparrow (\downarrow\Uparrow)$ refers to the polarization
of the electron parallel (antiparallel) to that of the target nucleon.

The structure functions $G_1$ and $G_2$ can be measured by taking the
difference of cross sections with electron and nucleon polarizations
parallel and antiparallel~\cite{HEY,LP}:
\begin{eqnarray}
\label{sigpol}
{ d^2\sigma^{\uparrow\Uparrow-\downarrow\Uparrow} \over d\Omega dE' }
&=& { 4 \alpha^2 \over Q^4 } { E' \over E }
    \left[ (E + E' \cos\theta) M G_1(x,Q^2) - Q^2 G_2(x,Q^2)
    \right]\ .
\end{eqnarray}

It will be convenient later to introduce dimensionless structure
functions $F_{1,2}$, $g_{1,2}$, defined as:
\begin{eqnarray}
M\ W_1 &=& F_1, \hspace*{1cm} \nu\ W_2\ =\ F_2	\nonumber\\
M^2 \nu\ G_1 &=& g_1, \hspace*{1cm} M \nu^2\ G_2\ =\ g_2\ .
\end{eqnarray}
As we will see in Section~3.2, some of these dimensionless structure
functions have very simple interpretations in terms of quark densities
in deep-inelastic scattering.

The structure functions describe scattering to final states whose
spectrum depends on the amount of energy and momentum transferred
to the target nucleon.
In Fig.~2 the differential cross section for unpolarized scattering
is plotted as a function of the invariant mass, $W$, of the hadronic
final state, where $W^2 = (p+q)^2 = M^2 + 2 M \nu - Q^2$.
At low energy, since $\nu > Q^2/2M$ only elastic scattering is
kinematically allowed (represented by the spike in Fig.~2 at $W = M$).
As the energy increases above the pion production threshold,
$W_{\rm th} = M+m_\pi$ (left vertical line in Fig.~2), inelastic
scattering to nucleon + multi-pion states can occur, as well as
excitation of nucleon resonances.
The first peak corresponds to the spin- and isospin-3/2 $\Delta$
resonance at $W=1232$~GeV.
The second peak is predominantly due to the negative parity partner
of the nucleon, the $S_{11}$ resonance, and the third to the $F_{15}$.
In the region between $W=W_{\rm th}$ and $W=2$~GeV many resonances
contribute, most of whose contributions are buried underneath the
background.
The vertical line in Fig.~2 at $W=2$~GeV corresponds to the approximate
boundary between the resonance and deep-inelastic scattering (DIS)
regions.
In the next Section we shall examine electron scattering to the
simplest final state, namely the elastic.

\begin{figure}
\label{sigexp}
\centering{\ \psfig{figure=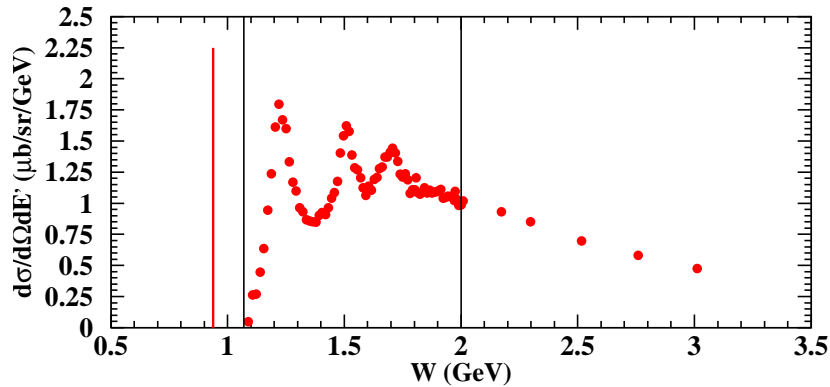,height=5cm}}
\caption{Electron--nucleon differential cross
	section~\protect\cite{IOANA} as a function of
	hadronic final state mass, $W$.}
\end{figure}

\subsection{Elastic Form Factors}

The most basic observables which reflect the composite nature of the
nucleon are its electromagnetic form factors.
Historically, the first indication that the nucleon is not elementary
came from measurements of the form factors in elastic electron--proton
scattering~\cite{FFP}.

The nucleon form factors are defined through matrix elements of the
electromagnetic current,
$J_{\mu} = \overline \psi \gamma_{\mu} \psi$, with $\psi$ the quark
field, as:
\begin{eqnarray}
\label{ffdef}
\langle N(P') | J_{\mu}(0) | N(P) \rangle
&=& \overline u(P')
\left( \gamma_{\mu} {\cal F}_1(Q^2)
     + {i\sigma_{\mu\nu} q^{\nu} \over 2 M} {\cal F}_2(Q^2)
\right) u(P)\ ,
\end{eqnarray}
where $P$ and $P'$ are the initial and final nucleon momenta,
and $q = P - P'$.
The functions ${\cal F}_1$ and ${\cal F}_2$ are the Dirac and Pauli
form factors, respectively.
In terms of ${\cal F}_1$ and ${\cal F}_2$ the Sachs electric and
magnetic form factors are defined as:
\begin{eqnarray}
G_E(Q^2) &=& {\cal F}_1(Q^2) - (Q^2/4M^2)\ {\cal F}_2(Q^2)\ ,	\\
G_M(Q^2) &=& {\cal F}_1(Q^2) + {\cal F}_2(Q^2)\ .
\end{eqnarray}
Squaring the amplitude in Eq.(\ref{ffdef}) and comparing with the cross
section in Eq.(\ref{sigdis}), one can write the structure functions for
elastic scattering in terms of the electromagnetic form factors:
\begin{eqnarray}
\label{sfelf1}
F_1^{\rm el}
&=& M \tau\
    G_M^2(Q^2)\ \delta\left( \nu - {Q^2 \over 2M} \right)\ ,	\\
\label{sfelf2}
F_2^{\rm el}
&=& { 2 M \tau \over 1 + \tau }
    \left( G_E^2(Q^2) + \tau G_M^2(Q^2) \right)\
    \delta\left( \nu - {Q^2 \over 2M} \right)\ ,
\end{eqnarray}
where $\tau = Q^2/4 M^2$.
For the spin-dependent structure functions one has:
\begin{eqnarray}
\label{sfelg1}
g_1^{\rm el}
&=& { M \tau \over 1 + \tau }
    G_M(Q^2) \left( G_E(Q^2) + \tau G_M(Q^2) \right)
    \delta\left( \nu - {Q^2 \over 2M} \right)\ ,		\\
\label{sfelg2}
g_2^{\rm el}  
&=& { M \tau^2 \over 1 + \tau }
    G_M(Q^2) \left( G_E(Q^2) - G_M(Q^2) \right)
    \delta\left( \nu - {Q^2 \over 2M} \right)\ .
\end{eqnarray}
At the quark level, the form factors can be decomposed as:
\begin{eqnarray}
G_{E,M}(Q^2) &=& \sum_q e_q\ G^q_{E,M}(Q^2)\ ,
\end{eqnarray}
so that contributions from specific quark flavors can be identified
by considering different hadrons.
For the proton one has:
\begin{eqnarray}
G^p_{E,M}
&=& {2 \over 3} G^u_{E,M}\
 -\ {1 \over 3} G^d_{E,M}\
 -\ {1 \over 3} G^s_{E,M}\ ,
\end{eqnarray}
while for the neutron (using isospin symmetry):
\begin{eqnarray}
G^n_{E,M}
&=& {2 \over 3} G^d_{E,M}\
 -\ {1 \over 3} G^u_{E,M}\
 -\ {1 \over 3} G^s_{E,M}\ ,
\end{eqnarray}
where $G^q_{E,M}$ by definition refers to the quark flavor
in the proton.

At low $Q^2$ the distance scales on which the electromagnetic scattering
takes place are large, so that the resolving power of the probe is only
sufficient to measure the static properties of the nucleon.
In this region the form factors reflect completely the non-perturbative,
long-distance structure of the nucleon.
At $Q^2=0$ the electric form factor is equal to the charge of the nucleon,
while the magnetic form factor gives the magnetic moment:
\begin{eqnarray}
G^N_E(0) &=& e_N\ ,\ \ \ \ G^N_M(0)\ =\ \mu_N\ ,
\end{eqnarray}
where $e_N = 1(0)$ for the proton (neutron), and in units of the Bohr
magneton, $\mu_p = 2.79 \mu_0$ and $\mu_n = -1.91 \mu_0$.

At small $Q^2$ away from zero, in a frame of reference where the energy
transfer to the nucleon is zero (namely, the Breit frame, $\nu=0$,
$Q^2 = \vec q^{\ 2}$), the electric and magnetic form factors measure
the Fourier transforms of the distributions of charge and magnetization,
$\rho_E$ and $\rho_M$, in the nucleon:
\begin{eqnarray}
G_{E,M}(Q^2) &=& \int d^3r\ e^{-i \vec q \cdot \vec r} \rho_{E,M}(r)\ .
\end{eqnarray}
Expanding the form factors about $Q^2=0$,
\begin{eqnarray}
G_{E,M}(Q^2)
&=& G_{E,M}(0) - { 1 \over 6 } Q^2 \langle r^2 \rangle_{E,M}
		+ {\cal O}(Q^4)\ ,
\end{eqnarray}
enables one to define the charge and magnetization radii of the nucleon
in terms of the slope of the form factors at $Q^2=0$:\
$\langle r^2 \rangle_{E,M}
 = -6 \left. dG_{E,M}/dQ^2 \right|_{Q^2=0}$.
Empirically, the form factors at low $Q^2$ are well described by a
dipole form,
\begin{eqnarray}
\label{GD}
G_E^p(Q^2)\ \approx\ { G_M^p(Q^2) \over \mu_p }\
	    \approx\ { G_M^n(Q^2) \over \mu_n }\
	    \approx\ G_D(Q^2)
	    &=& \left( { 1 \over 1 + Q^2/Q_0^2 } \right)^2,
\end{eqnarray}
where $Q_0^2 = 0.71$~GeV$^2$, in which case the electric and magnetic
r.m.s. radii of the proton, and the magnetic radius of the neutron,
are $\langle r^2 \rangle^{1/2} = 0.81$~fm.
Because the neutron has zero charge, the neutron electric form factor,
although non-zero, is very small.

The dipole form can be qualitatively understood within a vector meson
dominance picture, in which the photon at low $Q^2$ fluctuates into a
$J^P=1^-$ meson (such as the $\rho$), which then interacts with the
nucleon.
The $Q^2$ dependence of the form factor is then given by the vector
meson propagator, with the mass of the $\rho$ corresponding approximately
to $Q_0$.
However, while providing a reasonable approximation to the form factors
at low $Q^2$, deviations from the dipole form have been observed, and it
is important to understand the nature of the deviations at larger $Q^2$.

At the other extreme of asymptotically large $Q^2$, the elastic form
factors can be described in terms of perturbative QCD~\cite{LB}.
Here the short wavelength of the highly virtual photon enables the
quark substructure of the nucleon to be cleanly resolved.
By counting the minimal number of hard gluons exchanged between quarks,
the $Q^2$ behavior of the form factors is predicted from perturbative
QCD to be $G_{E,M}(Q^2) \sim 1/Q^4$ at large $Q^2$.
Just where the perturbative behavior sets in, however, is still an open
question which must be resolved experimentally.
Evidence from recent experiments at Jefferson Lab and elsewhere suggests
that non-perturbative effects are still quite important for $Q^2$ at
least $\approx 10$~GeV$^2$.

Understanding the transition from the low to high $Q^2$ regions is vital
not only for determining the onset of perturbative behavior.
Form factors in the transition region at intermediate $Q^2$ are very
sensitive to mechanisms of spin-flavor symmetry breaking, which cannot
be described within perturbation theory.
In Section~4 we give several examples where form factors reflect
important aspects of the non-perturbative structure of the nucleon.

\subsection{Deep-Inelastic Structure Functions}

Because of the $1/Q^4$ dependence in the elastic form factor in
Eq.(\ref{GD}), the elastic cross section dies out very rapidly with
$Q^2$.
It was therefore expected that the inelastic cross section would
behave in a similar fashion at large $Q^2$.
Contrary to the expectation, the observation~\cite{EARLYSLAC} in the
late 1960s that the inelastic structure function does not vanish with
$Q^2$, but remains approximately constant beyond $Q^2 \sim 2$~GeV$^2$,
provided the first evidence of point-like constituents of the
nucleon~\cite{FEYN69,BP} and led to the development of the parton model
and later QCD.

For inclusive scattering the sum in Eq.(\ref{wmunudef}) is taken over
all hadronic final states $X$.
Using the completeness relation $\sum_X | X \rangle \langle X | = 1$
and translational invariance, the hadronic tensor $W_{\mu\nu}$ can
be written:
\begin{eqnarray}
\label{wmunu}
W_{\mu\nu} &=& {1 \over 2\pi} \int d^4\xi\ e^{i q\cdot\xi}
             \langle N | J_{\mu}(\xi) J_{\nu}(0) | N \rangle\ ,
\end{eqnarray}
where $\xi$ is the space-time coordinate.
In the limit where $p \cdot q$ and $Q^2 \rightarrow \infty$, but the
ratio of these fixed (Bjorken limit), $W_{\mu\nu}$ receives its dominant
contributions from the light-cone region. 
This is clear if one writes the argument of the exponential in
light-cone coordinates,
\begin{eqnarray}
q \cdot \xi
&=& {q_+ \xi_- \over 2}\ +\ {q_- \xi_+ \over 2}\
			 -\ {\bf q}_\perp \cdot \xi_\perp\ ,
\end{eqnarray}
where $\xi_{\pm} = \xi_0 \pm \xi_z$.
In the target rest frame the photon momentum can be taken as\ 
$q_{\mu}$ = $\left( \nu; {\bf 0}_\perp, -\sqrt{\nu^2 + Q^2} \right)$
    $\simeq (\nu; {\bf 0}_\perp, - \nu - M x)$,
so that\ 
$q \cdot \xi = -M x (\xi_0 - \xi_z)/2\ $
 +\ $(2\nu + M x) (\xi_0 + \xi_z)/2$,
where in this frame $x=Q^2/2M\nu$.
The largest contributions to the integral are those for which the
exponent oscillates least, namely $q \cdot \xi \simeq 0$.
In the Bjorken limit $q \cdot \xi$ behaves like $\nu (\xi_0 + \xi_z)$,
so that only when $\xi_0 = -\xi_z$ will there be non-negligible
contributions to $W_{\mu\nu}$.
Therefore the DIS cross sections is controlled by the product of currents
$J_{\mu}(\xi) J_{\nu}(0)$ near the light cone, $\xi^2 \simeq 0$.

\subsubsection{Parton Model}

The connection between the deep-inelastic structure functions and the
quark structure of the nucleon was first provided by Feynman's parton
model~\cite{FEYN69}.
The hypothesis of the parton model is that the inelastic scattering
is described by incoherent {\em elastic} scattering from point-like,
spin 1/2 constituents (partons) in the nucleon.

The validity of the parton picture relies on the treatment of the
interactions of the virtual photon with the partons in the impulse
approximation.
The legitimacy of the impulse approximation rests on two assumptions:
(i) final state interactions can be neglected, and
(ii) the interaction time is less than the lifetime of the virtual state
of the nucleon as a sum of its on-shell constituents.

The first assumption seems reasonable since in DIS the energy
transferred to the parton is much greater than the binding energy,
so that the partons can be viewed as quasi-free.
The second assumption can be verified in the infinite momentum frame
(IMF), where the momentum of the nucleon is
$p_{\mu} = \left( p_z + M^2/2p_z;\ {\bf 0}_\perp,\ -p_z \right)$,
with $p_z \rightarrow \infty$ (or $v/c \rightarrow 1$),
in which case the photon four-momentum is
$q_{\mu} = \left( -xp_z (1-M^2 x^2/Q^2) + M \nu/2p_z;\ {\bf 0}_\perp,\
\right.$
          $\left.  xp_z (1-M^2 x^2/Q^2) + M \nu/2p_z \right)$.
Since the dominant contributions to $W_{\mu\nu}$ are those
for which $q \cdot \xi \ll 1$ and $\xi_0 \simeq -\xi_z$, one has
$q \cdot \xi \simeq 2 M \nu \xi_0/p_z$,
so that the interaction time is $\sim \xi_0 \leq p_z/2 M \nu$.
The lifetime of the virtual state can be obtained by observing that
the energy of the virtual nucleon state consisting of on-shell partons
with momenta $x_i p_z$ and mass $m_i$ is
$\approx \sum_i (x_i p_z + m_i^2/2 x_i p_z)$,
so that the difference between the energies of the virtual and on-shell
nucleons is $\approx (\sum_i m_i^2/x_i - M^2)/2p_z$.
Therefore the lifetime of this virtual state is proportional to $p_z$,
and in the Bjorken limit the ratio of interaction time to virtual state
lifetime $\sim 1/\nu \rightarrow 0$.

The parton picture is then of a quark with momentum fraction $x_i$
absorbing a photon with $x_i = x$, since
$\delta ( (q + x_i p)^2 ) \rightarrow \delta ( x - x_i ) / 2 p\cdot q$.
One can then relate the structure functions to the parton densities
(quark and antiquark momentum distribution functions) in the nucleon
as~\cite{FEYN69,BP}:
\begin{eqnarray}
\label{partonf1}
F_2(x)
&=& \sum_q\ e_q^2\ x (q(x) + \bar q(x))\ =\ 2 x F_1(x)\ ,	\\
\label{partong1}
g_1(x)
&=& {1 \over 2} \sum_q\ e_q^2\ x (\Delta q(x) + \Delta \bar q(x))\ ,
\end{eqnarray}
where $q(x) = q^{\uparrow}(x) + q^{\downarrow}(x)$
and $\Delta q(x) = q^{\uparrow}(x) - q^{\downarrow}(x)$
are the spin-averaged and spin-dependent quark densities.
The consequence of point-like partons is the non-vanishing of the
inelastic structure functions at large momentum transfers, since the
structure functions in Eqs.(\ref{partonf1}) and (\ref{partong1}) are
independent of $Q^2$.

Although providing a simple, intuitive language in which to interpret
the qualitative features of the deep-inelastic data, the parton model
is not a field theory.
The formal basis for the parton model is provided by the operator
product expansion and the renormalization group equations in QCD,
which actually gives rise to small violations of scaling through QCD
radiative corrections.

\subsubsection{Operator Product Expansion}

In quantum field theory products of operators at the same space-time
point (composite operators) are not well defined~\cite{MUTA}.
The short distance operator product expansion (OPE) of Wilson~\cite{OPE},
in which the composite operators are expanded in a series of finite local
operators multiplied by singular coefficient functions, provides a way of
obtaining meaningful results.

Because deep-inelastic scattering probes the $\xi^2 \sim 0$ region,
rather than the $\xi \sim 0$, one needs an expansion of the product of
currents in Eq.(\ref{wmunu}) which is valid near the light-cone
(this is because at short distances $Q \rightarrow \infty$ and
$p\cdot q/Q^2 \rightarrow 0$, while in DIS the light-cone region
corresponds to the Bjorken limit, $Q^2 \rightarrow \infty$ and
$p\cdot q/Q^2 = O(1)$).
The general form of the light-cone operator product expansion
is~\cite{OPE}:
\begin{eqnarray}
J(\xi) J(0)
&\sim&  \sum_{i,N} C^N_i(\xi^2)\ \xi_{\mu_1} \cdots \xi_{\mu_N}\
                   {\cal O}_i^{\mu_1 \cdots \mu_N}(0)\ ,
\end{eqnarray}
where the sum is over different types of operators with spin $N$
(i.e. those what transform as tensors of rank $N$ under Lorentz
transformations).
In DIS the spin-$N$ operators ${\cal O}_i^{\mu_1 \cdots \mu_N}$
represent the soft, non-perturbative, physics, while the coefficient
functions $C^N_i$ describe the hard photon--quark interaction,
and are calculable within perturbative QCD.

It is useful to categorize the operators according to their flavor
properties, namely those that are invariant under SU($N_f$) flavor
transformations (singlet) and those that are not (non-singlet).
For unpolarized scattering (the extension to spin-dependent scattering
is straightforward) the operators must be completely symmetric with
respect to interchange of indices $\mu_1 \cdots \mu_N$, so that one can
construct at most 3 kinds of composite operators~\cite{BURAS,MUTA}.
The non-singlet operators must be bilinear in the quark fields:
\begin{eqnarray}
\label{opNS}
{\cal O}_{NS}^{\mu_1 \cdots \mu_N}
&=& {i^{N-1} \over 2\ N!}\
  \overline\psi \left( \gamma^{\mu_1} D^{\mu_2} \cdots D^{\mu_N}\
               +\ \mu_i \mu_j\ {\rm permutations}
           \right)\ \vec{\lambda}\ \psi\ ,
\end{eqnarray}
where $\vec{\lambda}$ are the eight Gell-Mann matrices 
of the flavor SU($N_f$) group.
The singlet operators are:
\begin{eqnarray}
\label{opS}
{\cal O}_{\psi}^{\mu_1 \cdots \mu_N}
&=& {i^{N-1} \over N!}
    \overline\psi \left( \gamma^{\mu_1} D^{\mu_2} \cdots D^{\mu_N}\
                 +\ \mu_i \mu_j\ {\rm permutations}
             \right)\ \psi\ ,					\\
{\cal O}_G^{\mu_1 \cdots \mu_N}
&=& {i^{N-2} \over 2\ N!}
    \left( F^{\mu_1\alpha}
           D^{\mu_2} \cdots D^{\mu_{N-1}}
           F^{\mu_N}_{\alpha}\
        +\ \mu_i \mu_j\ {\rm permutations}
   \right),
\label{opG}
\end{eqnarray}
corresponding to the quark and gluon fields, respectively,
and color indices have been suppressed.

Equations (\ref{opNS})-(\ref{opG}) represent operators with the lowest
`twist', defined as the difference between the mass dimension and the
spin, $N$, of the operator.
Whereas the leading twist terms involve free quark fields, operators
with higher twist involve both quark--gluon interactions~\cite{JAF83},
for example $\overline\psi \widetilde F^{\mu\nu} \gamma_\nu \psi$,
and are suppressed by powers of $1/Q^2$.

The matrix elements of the operators contain information about
the long-distance, non-perturbative structure of the nucleon.
They can in general be written as:
\begin{eqnarray}
\langle N(p) | {\cal O}_i^{\mu_1 \cdots \mu_N} | N(p) \rangle
&=& {\cal A}_i^N\  p^{\mu_1} \cdots p^{\mu_N}\
 -\ (g^{\mu_i\mu_j}\ {\rm terms})\ ,
\end{eqnarray}
where ${\cal A}_i^N$ represents the soft physics, and the `trace terms'
containing the $g^{\mu_i\mu_j}$ are necessary to ensure that the matrix
elements are traceless (i.e. so that the composite operator has definite
spin, $N$).
When contracted with the $q_{\mu_i} q_{\mu_j}$ these give rise to terms
that contain smaller powers of $\nu^2$ (i.e. $Q^2 = O(\nu)$ instead of
$(p\cdot q)^2 = O(\nu^2)$).

The OPE analysis allows one to factorize the moments of the structure
functions into short and long distance contributions, where the latter
are target-dependent (and $Q^2$-independent).
For the $F_2$ structure function, for example, one has:
\begin{eqnarray}
\label{momdef}
M_2^N(Q^2) \equiv \int_0^1 dx\ x^{N-2}\ F_2(x,Q^2)
&=& \sum_i C^N_i(Q^2)\ {\cal A}^N_i\ .
\end{eqnarray}
The target-independent (and $Q^2$-dependent) coefficient functions
$C_i^N$ can be calculated at a given order in perturbation theory
directly from the renormalization group equations, which will introduce
logarithmic $Q^2$ violations of scaling compared with the simple parton
model.

The structure function can be obtained from the moments via the
inverse Mellin transform:
\begin{eqnarray}
\label{mellin}
F_2(x,Q^2) &=& {1 \over 2\pi i} \int_{N_0-i\infty}^{N_0+i\infty}
                    dN\ x^{1-N}\ M_2^N(Q^2)\ ,
\end{eqnarray}
by fixing the contour of integration to lie to the right of all
singularities of $M_2^N(Q^2)$ in the complex-$N$ plane.
In terms of the quark and gluon distributions, $F_2$ is given
by a convolution of the parton densities with the coefficient
functions~\cite{BURAS,MUTA,WEIGL} describing the hard photon--parton
interaction:
\begin{eqnarray}
\label{f2con}
F_2^p(x,Q^2)
&=& {1 \over 6}
    \int_x^1\ {dy \over y}
    \left( C_{NS}\left(x,Q^2\right)\
           x q_{NS}\left( {x\over y},Q^2 \right)
    \right.                                     \nonumber\\
& & \hspace*{-1.5cm}
    \left.
        +\ {5 \over 3} C_q \left(x,Q^2\right)\
           x \Sigma\left( {x \over y},Q^2\right)
        +\ {5 \over 3} C_G \left(x,Q^2\right)\
           x G\left( {x \over y},Q^2\right)
    \right)\ ,
\end{eqnarray}
where $q_{NS}(x,Q^2) = (u + \bar u - d - \bar d - s - \bar s)(x,Q^2)$
is the flavor non-singlet combination (for three flavors), while the
singlet combination $\Sigma(x,Q^2) = \sum_q (q + \bar q)(x,Q^2)$,
and $G(x,Q^2)$ is the gluon distribution.
The gluon coefficient $C_G$ enters only at order $\alpha_s$
(since the photon can only couple to the gluon via a quark loop).

The challenge to understanding the quark and gluon structure of
hadrons is to calculate the soft matrix elements ${\cal A}^N_i$
in Eq.(\ref{momdef}), or equivalently the parton distributions
in Eq.(\ref{f2con}), from QCD.
At present this can be only be done numerically through lattice QCD,
or in QCD-inspired quark models of the nucleon.
Considerable progress has been made over the last two decades in
attempting to establish a connection between the high energy parton
picture of DIS on the one hand, and the valence quark models at low
energy on the other~$^{24-26}$.
The underlying philosophy~\cite{JAFROS,PARPETR} has been that if the
nucleon behaves like three valence quarks at some low momentum scale
$\sim \mu^2$, a purely valence quark model may yield reliable twist-two
structure functions.
Comparison with experiment at DIS scales, where a description in terms
of valence quarks will no longer be accurate, can then be made by
evolving the structure function to higher $Q^2$ via the renormalization
group equations.

\subsubsection{Renormalization Group Equations}

In an interacting field theory like QCD quantities such as coupling
constants, masses, as well as wave functions (operators), must be
renormalized.
The renormalization procedure introduces an arbitrary renormalization
scale, $\mu^2$, into the theory, although of course the physics itself
cannot depend on $\mu^2$.

In the following we will consider the renormalization of the non-singlet
operators corresponding to the $F_2$ structure function.
The generalization to singlet operators in straightforward, although
one needs to take into account mixing between the singlet quark and
gluon operators~\cite{MUTA}.
If the unrenormalized matrix elements of the OPE are independent of
the renormalization scale $\mu$, then
\begin{eqnarray}
\label{operen}
{d \over d\mu} \langle N(p) | J_{\mu}(\xi) J_{\nu}(0) | N(p) \rangle
&=& 0\ .
\end{eqnarray}
Defining the wave function renormalization of the spin-$N$ non-singlet
operator by ${\cal O}_{\rm bare}^N = Z^N\ {\cal O}^N_{\rm ren}$, where
$Z^N$ is the renormalization constant, Eq.(\ref{operen}) can be rewritten
as:
\begin{eqnarray}
\label{rge}
\left( \mu {\partial \over \partial\mu}
     + \beta(g) {\partial \over \partial g}
     - \gamma^N
\right)\ C^N(Q^2/\mu,g) &=& 0\ ,
\end{eqnarray}
for each spin $N$.
This is the well-known renormalization group equation for the coefficient
functions.
In Eq.(\ref{rge}), the strong coupling constant $g$ is renormalized at
the scale $\mu^2$, and $\gamma^N$ is the anomalous dimension of the
twist-two operator ${\cal O}^N$:
\begin{eqnarray}
\label{anomdim}
\gamma^N
&=& \mu {\partial \over \partial \mu} (\ln Z^N)\ ,
\end{eqnarray}
and the $\beta$-function is given by:
\begin{eqnarray}
\beta(g) &=& \mu\ { \partial g \over \partial \mu }\ .
\end{eqnarray}
%
%
The solution to Eq.(\ref{rge}) is:
\begin{eqnarray}
C^N(Q^2/\mu^2,g^2)
&=& C^N(1,\bar{g}^2)
\exp \left[ - \int_{\bar g(\mu^2)}^{\bar g(Q^2)}
              dg'\ {\gamma^N(g') \over \beta(g')}
     \right]\ ,
\end{eqnarray}
where $\bar g$ is the effective (running) coupling constant,
defined by $d\bar{g}^2 / dt\ =\ \bar{g}\ \beta(\bar{g})$,
with $t = \ln(Q^2/\mu^2)$ and $\bar{g}(t=0)\ =\ g$.
The coefficients $C^N$, anomalous dimension $\gamma^N(g)$ and the
$\beta$-function can all be calculated in perturbation theory by
expanding in powers of the coupling, $g$:
\begin{eqnarray}
\gamma^N(g)
&=& \gamma^{(0) N}\ { g^2 \over 16\pi^2 }\ +\ O(g^2)\ , \nonumber\\
\beta(g)
&=& -\beta_0\ { g^3 \over 16\pi^2 }\ +\ O(g^5)\ ,		\\
C^N(1,\bar{g}^2)
&=& C^{(0) N}\ +\ O(g^2)\ .				\nonumber
\end{eqnarray}
It is then straightforward to derive the equation governing the $Q^2$
evolution of the non-singlet moments of the structure functions, which
to lowest order in $g$ is~\cite{GP74,GW74}:
\begin{eqnarray}
\label{momevol}
{\cal Q}^N(Q^2) &=& \int_0^1 dx x^{N-1} q_{NS}(x)\ =\
\left[ { \alpha_s(Q^2) \over \alpha_s(\mu^2) }
\right]^{ \gamma^{(0) N} / 2 \beta_0 }
{\cal Q}^N(\mu^2)\ ,
\end{eqnarray}
where the lowest order non-singlet anomalous dimension is:
\begin{eqnarray}
\gamma^{(0) N}
&=& {8 \over 3} \left( 4 \sum_{j=1}^N {1\over j} - 3 - {2\over N(N+1)}
\right)\ ,
\end{eqnarray}
and $\beta_0 = 11-2 N_f/3$ for $N_f$ active flavors in the evolution.
The strong coupling constant has been rewritten in Eq.(\ref{momevol}) as:
\begin{eqnarray}
\alpha_s(Q^2)
\equiv {\bar{g}^2(Q^2) \over 4 \pi}
&=& { 4\pi \over \beta_0 \ln (Q^2/\Lambda_{\rm QCD}^2) }\ ,
\end{eqnarray}
by putting the arbitrariness of the renormalization scale into the
parameter $\Lambda_{\rm QCD}$, known as the QCD scale parameter,
$\ln \Lambda_{\em QCD}^2
= \ln \mu^2 - 16\pi^2 / (\beta_0\ \bar{g}^2(\mu^2))$.
Once the moments of the structure function or parton distribution
are known at $\mu^2$, Eq.(\ref{momevol}) can be used to give the
moments at any other value of $Q^2$.

An intuitive and mathematically equivalent picture for this $Q^2$
evolution is provided by the DGLAP evolution equations~\cite{DGLAP}.
For the non-singlet quark distribution one has:
\begin{eqnarray}
\label{intdif}
{ d q_{NS}(x,Q^2) \over dt }
&=& { \alpha_s(Q^2) \over 2\pi }
\int_x^1 {dy \over y}\ q_{NS}(y,t)\ P_{qq}(x/y)\ ,
\end{eqnarray}
where $P_{qq}(x/y)$ is the $q \rightarrow q + g$ splitting function,
which gives the probability of finding a quark with momentum fraction
$x$ inside a parent quark with momentum fraction $y$, after it has
radiated a gluon.
The splitting function is related to the anomalous dimension by\ 
$\gamma^N \sim \int dz z^{N-1} P(z)$.
The generalization to singlet evolution is again straightforward,
but involves a set of coupled quark and gluon
equations~\cite{BURAS,MUTA,WEIGL}.

The physical interpretation of the DGLAP equations is that as $Q^2$
increases the quarks radiate more and more gluons, which subsequently
split into quark and antiquark pairs, which themselves then radiate
more gluons, and so on.
In this manner the quark-antiquark sea can be generated from a pure
valence component.
This process modifies the population density of quarks as a function
of $x$, so that the momentum carried by quarks is no longer a static
property of the nucleon, but now depends on the resolving power of
the probe, $Q^2$.
In general, the larger the $Q^2$, the better the resolution, and the
more substructure seen in the hadron.
It is a remarkable success of perturbative QCD that it can provide
a quantitative description of the scaling violations of structure
functions~$^{32-34}$ for a large range of $x$ and over many orders of
magnitude of $Q^2$.

\subsection{Semi-Inclusive Scattering}

Inclusive electron--nucleon scattering is a well-established tool
which has been used to study nucleon structure for many years.
Somewhat less exploited, but potentially more powerful,
is semi-inclusive scattering~\cite{SEMI}, in which a specific hadron,
$h$, is observed in coincidence with the scattered electron,\
$e N \rightarrow e' h X$.
This process offers considerably more freedom to explore the individual
quark content of the nucleon than is possible through inclusive
scattering~\cite{EPIC}.

A central assumption in semi-inclusive DIS is that at high energy the
interaction and production processes factorize.
Namely, the interaction of the virtual photon with a parton takes place
on a much shorter space-time scale than the fragmentation of the struck
quark, and the spectator quarks, into final state hadrons.
Furthermore, the hadronic products of the scattered quark (in the current
fragmentation region, along the direction of the current in the
photon--nucleon center of mass frame) should be clearly separated from
the hadronic remnants of the target (in the target fragmentation region).

The cleanest way to study fragmentation is in the current fragmentation
region, where the scattered quark fragments into hadrons by picking up
$q \bar q$ pairs from the vacuum.
The production of a specific final state hadron, $h$, is parameterized
by a fragmentation function, $D_q^h(z)$, which gives the probability
of quark $q$ fragmenting into a hadron $h$ with a fraction $z$ of the
quark's (or, at high energy, the photon's) center of mass energy.
Because it requires only a single $q\bar q$ pair, the leading hadrons
in this region are predominantly mesons.
At large $z$, where the knocked out quark is most likely to be contained
in the produced meson, one can obtain direct information on the momentum
distribution of the scattered quark in the target.
At small $z$ this information becomes diluted by additional $q\bar q$
pairs from the vacuum which contribute to secondary fragmentation.

In the QCD-improved parton model the number of hadrons, $h$, produced
at a given $x$, $z$ and $Q^2$ can be written (in leading order) as:
\begin{eqnarray}
\label{sidis}
N^h(x,z,Q^2)
&\propto& \sum_q\ e_q^2\ q(x,Q^2)\ D_q^h(z,Q^2)\ .
\end{eqnarray}
Although factorization of the $x$ and $z$ dependence is generally true
only at high energy, recent data from HERMES~\cite{HERMES_DU} suggests
that at $\nu \sim$ 10--20~GeV the fragmentation functions are still
independent of $x$, and agree with previous measurements by the
EMC~\cite{EMCFRAG} at somewhat larger energies.
Where the factorization hypothesis breaks down is not known, and
the proposed semi-inclusive program at an energy upgraded CEBAF at
Jefferson Lab, with $\nu$ typically $\sim$ 5--10~GeV, will test the
limits of the parton interpretation of meson electroproduction.

\subsection{Off-Forward Parton Distributions}

The nucleon's deep-inelastic structure functions and elastic form factors
parameterize fundamental information about its quark substructure.
Both reflect dynamics of the internal quark wave functions describing   
the same physical ground state, albeit in different kinematic regions.
An example of how in certain cases these are closely related is provided
by the phenomenon of quark--hadron duality (Section~4.1).

Recently it has been realized that form factors and structure functions
can be simultaneously embedded within the general framework of
off-forward (sometimes also referred to as non-forward, or skewed)
parton distributions~\cite{OFPD,NFPD}.
The off-forward parton distributions (OFPDs) generalize and interpolate
between the ordinary parton distributions in deep-inelastic scattering
and the elastic form factors.

\begin{figure}
\label{fig:ofpd}
\centering{\ \psfig{figure=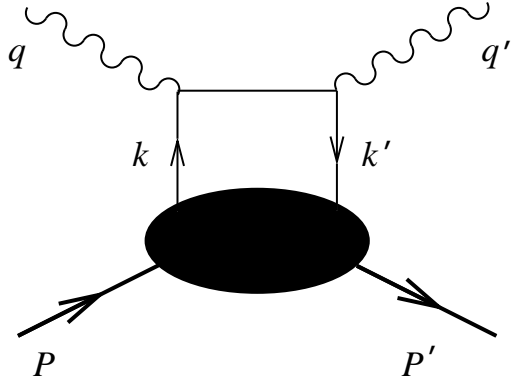,height=5cm}}
\caption{Leading twist off-forward parton distribution, as seen in
	deeply-virtual Compton scattering.  The nucleon, parton and
	photon initial (final) momenta are labeled $P$ ($P'$),
	$k$ ($k'$) and $q$ ($q'$), respectively.
	Deep-inelastic scattering corresponds to $q=q'$, $P=P'$.}
\end{figure}

As illustrated in Fig.~3, the OFPD is the amplitude (in the infinite
momentum frame) to remove a parton with momentum $k_\mu$ from a nucleon
of momentum $P_\mu$, and insert it back into the nucleon with momentum
$k'_\mu = k_\mu + P'_\mu - P_\mu$, where $P'_\mu$ is the final state
nucleon momentum.
The simplest physical process in which the OFPD can be measured is
deeply-virtual Compton scattering~\cite{OFPD} (DVCS).

For the spin-averaged case, the OFPDs can be defined as matrix elements
of bilocal operators
$\overline \psi(-\lambda n/2) {\cal L} \gamma^{\mu} \psi(\lambda n/2)$,
where $\lambda$ is a scalar parameter, 
and $n_\mu$ is a light-like vector proportional to $(1;0,0,-1)$.
The gauge link ${\cal L}$, which is along a straight line segment
extending from one quark field to the other, makes the bilocal operator
gauge invariant.
In the light-like gauge $A^\mu n_\mu = 0$, so that the gauge link
is unity.
The most general expression for the leading contributions at large $Q^2$
can then be written~\cite{OFPD}:
\begin{eqnarray}
\label{ofpd}
\int {d\lambda \over 2\pi} e^{i\lambda x}
\langle P'|\overline\psi(-\lambda n/2)\gamma^\mu \psi(\lambda n/2) |P
\rangle
&=& H(x,\xi,t)\
    \overline u(P')\gamma^\mu u(P) \nonumber \\
& & \hspace*{-3cm} +\ E(x,\xi,t)\ 
    \overline u(P') i\sigma^{\mu\nu} {(P'_\nu-P_\nu) \over 2M} u(P)
    + \cdots  \ ,
\end{eqnarray}
where $t = (P'-P)^2$ and $\xi = -n\cdot (P'-P)$,
with $u(P)$ the nucleon spinor, and the dots ($\cdots$) denote
higher-twist contributions.   
A similar expression can be derived for the axial vector current.
The structures in Eqs.(\ref{ofpd}) are identical to those in the
definition of the nucleon's elastic form factors, Eq.(\ref{ffdef}).
The chiral-even distribution, $H$, survives in the forward limit in which
the nucleon helicity is conserved, while the chiral-odd distribution, $E$,
arises from the nucleon helicity flip associated with a finite momentum
transfer~\cite{OFPD}.

The off-forward parton distributions display characteristics of both
the forward parton distributions and nucleon form factors.
In the limit of $P'_\mu \rightarrow P_\mu$, one finds~\cite{OFPD}:
\begin{equation}
H(x,0,0) = q(x)\ ,
\end{equation}
where $q(x)$ is the forward quark distribution, defined through similar
light-cone correlations~\cite{JJ} (the dependence on the scale, $Q^2$,
of $H$ and $q$ is suppressed).
On the other hand, the first moment of the off-forward distributions
are related to the nucleon form factors by the following sum
rules~\cite{OFPD}:
\begin{eqnarray}
\label{sumH}
\int^1_{-1} dx H(x,\xi,t)
&=& {1 \over 1+\tau} (G_E(t)+\tau G_M(t))\ ,		\\
\label{sumE}
\int^1_{-1} dx E(x,\xi,t)
&=& {1 \over 1+\tau} (G_M(t)-G_E(t))\ ,
\end{eqnarray}
where the $x$-integrated distributions are in fact independent of $\xi$.
These sum rules provide important constraints on any model calculation
of the OFPDs~\cite{OFPDBAG}.

Higher moments of the OFPDs can also be related to matrix elements of the
QCD energy-momentum tensor.
Because the form factors of the energy-momentum tensor contain
information about the quark and gluon contributions to the nucleon
angular momentum, the OFPDs can therefore provide information on the
fraction of the nucleon spin carried by quarks and gluons, which has
been a subject of intense interest now for more than a decade
(see Section~4.4).

Having introduced the tools necessary to study nucleon substructure
in inclusive and exclusive reactions, in the next Section we examine
more closely the dependence on flavor and spin of the quark momentum
distributions.

\section{Flavor and Spin Content of the Nucleon}

The distribution of quarks in the nucleon is perhaps the most
fundamental problem in hadron physics.
Knowing the total structure functions and form factors as a function
of $x$ and $Q^2$ is important for determining the scaling properties
and global characteristics of the nucleon, however, understanding the
relative contributions from different quark flavors gives us deeper
understanding of the nucleon's internal structure and dynamics.
In this Section we first explore the flavor dependence of the valence
quark distributions at large $x$, then discuss the flavor structure of
the quark-antiquark sea.
Finally, we review several topical issues concerning the spin structure
of the nucleon.

\subsection{Valence Quarks}

Much of the emphasis in recent years has been placed on exploring the
region of small Bjorken-$x$ at high-energy colliders such as HERA.
Delving into the small-$x$ region is necessary in order to determine
integrals of structure functions and minimize $x \rightarrow 0$
extrapolation errors when testing various integral sum rules.

At small $x$ ($x < 0.2$) most of the strength of the structure function
is due to the quark--antiquark sea generated through perturbative gluon
radiation and subsequent splitting into quark--antiquark pairs,
$g \rightarrow \bar q q$.
Genuine non-perturbative effects associated with the nucleon ground
state structure are therefore more difficult to disentangle from the
perturbative background.

Valence quark distributions, on the other hand, reflect essentially
long-distance, or non-perturbative, aspects of nucleon structure,
and can be more directly connected with low energy
phenomenology~$^{24-26}$ associated with form factors and
the nucleon's static properties.
After many years of structure function measurements over a range of
energies and kinematical conditions, the valence quark structure has
for some time now been thought to be understood.
However, there is one major exception --- the deep valence region,
at $x > 0.7$.

Knowledge of quark distributions at large $x$ is essential for a number
of reasons.
Not least of these is the necessity of understanding backgrounds in
collider experiments, such as in searches for new physics beyond the
standard model~\cite{CTEQ_LX}.
Furthermore, the behavior of the ratio of valence $d$ to $u$ quark
distributions in the limit $x \to 1$ provides a critical test of the
mechanism of spin-flavor symmetry breaking in the nucleon, and a test
of the onset of perturbative behavior in large-$x$ structure
functions~\cite{MTNP}.

\subsubsection{SU(6) Symmetry Breaking and the $d/u$ Ratio}

The precise mechanism for the breaking of the spin-flavor SU(6) symmetry
is a basic question for hadron structure physics.
In a world of exact SU(6) symmetry, the wave function of a proton,
polarized say in the $+z$ direction, would be given by~\cite{CLOSE79}:
\begin{eqnarray}
\label{pwfn}
p\uparrow
&=& {1 \over \sqrt{2}}  u\uparrow (ud)_{S=0}\
 +\ {1 \over \sqrt{18}} u\uparrow (ud)_{S=1}\
 -\ {1 \over 3}         u\downarrow (ud)_{S=1}\ \nonumber \\
& &
 -\ {1 \over 3}         d\uparrow (uu)_{S=1}\
 -\ {\sqrt{2} \over 3}  d\downarrow (uu)_{S=1}\ ,
\end{eqnarray}
where the subscript $S$ denotes the total spin of the two-quark
component.
In this limit, apart from charge and isospin, the $u$ and $d$ quarks
in the proton would be identical, and the nucleon and $\Delta$ would,
for example, be degenerate in mass.
In deep-inelastic scattering, exact SU(6) symmetry would be
manifested in equivalent shapes for the valence quark
distributions of the proton, which would be related simply by
$u_{\rm val}(x)=2 d_{\rm val}(x)$ for all $x$.
{}From Eq.(\ref{partonf1}), for the neutron to proton structure function
ratio this would imply:
\begin{eqnarray}
{ F_2^n \over F_2^p }
&=& {2 \over 3}\ \ \ \ \ \ {\rm [SU(6)\ symmetry]}.
\end{eqnarray}

In nature spin-flavor SU(6) symmetry is, of course, broken.
The nucleon and $\Delta$ masses are split by some 300 MeV.
Furthermore, it is known that the $d$ quark distribution in DIS is
considerably softer than the $u$ quark distribution, with the
neutron/proton ratio deviating at large $x$ from the SU(6) expectation.
The correlation between the mass splitting in the \underline{{\bf 56}}
baryons and the large-$x$ behavior of $F_2^n/F_2^p$ was observed some
time ago~$^{46-48}$.
Based on phenomenological~\cite{CLOSE73} and Regge~\cite{CARLITZ}
arguments, the breaking of the symmetry in Eq.(\ref{pwfn}) was argued
to arise from a suppression of the `diquark' configurations having
$S=1$ relative to the $S=0$ configuration, namely:
\begin{eqnarray}
\label{S0dom}
(qq)_{S=0} &\gg& (qq)_{S=1}, \ \ \ \ \ x \rightarrow 1\ .
\end{eqnarray}
Such a suppression is in fact quite natural~\cite{CT,ISGUR_V} if one
observes that whatever mechanism leads to the observed $N-\Delta$
splitting (e.g. color-magnetic force, instanton-induced interaction,
pion exchange), it necessarily acts to produce a mass splitting between
the two possible spin states of the two quarks which act as spectators
to the hard collision, $(qq)_S$, with the $S=1$ state heavier than the
$S=0$ state by some 200~MeV.
{}From Eq.(\ref{pwfn}), a dominant scalar valence diquark component
of the proton suggests that in the $x \rightarrow 1$ limit $F_2^p$
is essentially given by a single quark distribution (i.e. the $u$),
in which case:
\begin{eqnarray}
{ F_2^n \over F_2^p }
&\rightarrow& { 1 \over 4 }\ , \ \ \ \ \
{ d \over u } \rightarrow 0\ \ \ \ \
[S=0\ {\rm dominance}]\ .
\end{eqnarray}
This expectation has, in fact, been built into most phenomenological
fits to the parton distribution data~$^{32-34}$.

An alternative suggestion, based on perturbative QCD, was originally
formulated by Farrar and Jackson~\cite{FJ}.
There it was argued that the exchange of longitudinal gluons, which are
the only type permitted when the spins of the two quarks in $(qq)_S$ are
aligned, would introduce a factor $(1-x)^{1/2}$ into the Compton amplitude
--- in comparison with the exchange of a transverse gluon between quarks
with spins anti-aligned.
In this approach the relevant component of the proton valence wave
function at large $x$ is that associated with states in which the total
`diquark' spin {\em projection}, $S_z$, is zero:
\begin{eqnarray}
(qq)_{S_z=0} &\gg& (qq)_{S_z=1}, \ \ \ \ \ x \rightarrow 1\ .
\end{eqnarray}
Consequently, scattering from a quark polarized in the opposite direction
to the proton polarization is suppressed by a factor $(1-x)$ relative to
the helicity-aligned configuration.

This is related to the treatment based on counting rules where the
large-$x$ behavior of the parton distribution for a quark polarized
parallel ($\Delta S_z=1$) or antiparallel ($\Delta~S_z~=~0$) to the
proton helicity is given by\
\mbox{$q^{\uparrow\downarrow}(x) = (1-x)^{2n - 1 + \Delta S_z}$},
where $n$ is the minimum number of non-interacting quarks
(equal to 2 for the valence quark distributions).
In the $x \rightarrow 1$ limit one therefore predicts:
\begin{eqnarray}
{ F_2^n \over F_2^p }
&\rightarrow& {3 \over 7}\ , \ \ \ \ \
{ d \over u } \rightarrow { 1 \over 5 }\ \ \ \ \
[S_z=0\ {\rm dominance}]\ .
\end{eqnarray}
Similar predictions can be made for the ratios of polarized quark
distributions at large $x$ (see Section~4.4).

The biggest obstacle to an unambiguous determination of $d/u$ at
large~$x$ is the absence of free neutron targets.
In practice essentially all information about the structure of the
neutron is extracted from light nuclei, such as the deuteron.
The deuteron cross sections must however be corrected for nuclear
effects in the structure function, which can become quite
significant~\cite{SMEAR} at large $x$.
In particular, whether one corrects for Fermi motion only, or in
addition for binding and nucleon off-shell effects~\cite{MTNP},
the extracted $F_2^n/F_2^p$ ratio can differ by up to $\sim 50\%$
for beyond $x \sim 0.5$.

A number of suggestions have been made how to avoid the nuclear
contamination problem~$^{53-57}$.
One of the more straightforward ones is to measure relative yields of
$\pi^+$ and $\pi^-$ mesons in semi-inclusive scattering from protons
in the current fragmentation region~\cite{SEMIPI}.
At large $z$ ($z$ being the energy of the pion relative to the photon)
the $u$ quark fragments primarily into a $\pi^+$, while a $d$ fragments
into a $\pi^-$, so that at large $x$ and $z$ one has a direct measure
of $d/u$.

{}From Eq.(\ref{sidis}) the number of charged pions produced from a
proton target per interval of $x$ and $z$ is at leading order in QCD
given by~\cite{CLOSE79}:
\begin{eqnarray}
N_p^{\pi^+}(x,z,Q^2) &\sim& 4 u(x,Q^2)\ D(z,Q^2)\
			   +\ d(x,Q^2)\ \overline D(z,Q^2)\ ,	\\
N_p^{\pi^-}(x,z,Q^2 &\sim& 4 u(x,Q^2)\ \overline D(z,Q^2)\
			   +\ d(x,Q^2)\ D(z,Q^2)\ ,
\end{eqnarray}
where $D = D_u^{\pi^+} = D_d^{\pi^-}$ is the leading
fragmentation function (assuming isospin symmetry),
and $\overline D(z) = D_d^{\pi^+} = D_u^{\pi^-}$
is the non-leading fragmentation function.
Taking the ratio of these one finds:
\begin{eqnarray}
\label{Rpi}
R^{\pi}(x,z,Q^2)\ =\
{ N_p^{\pi^-} \over N_p^{\pi^+} }
&=& { 4 \overline D/D + d/u \over 4 + d/u \cdot \overline D/D }\ .
\end{eqnarray}
In the limit $z \rightarrow 1$, the leading fragmentation function
dominates, $D \gg \overline D$, and the ratio
$R^{\pi} \rightarrow (1/4) d/u$.

In the realistic case of smaller $z$, the $\overline D/D$ term in $R^\pi$
contaminates the yield of fast pions originating from struck primary
quarks, diluting the cross section with pions produced from secondary
fragmentation by picking up extra $q\bar q$ pairs from the vacuum.
Nevertheless, one can estimate the yields of pions using the empirical
fragmentation functions measured by the HERMES
Collaboration~\cite{HERMES_DU} and the EMC~\cite{EMCFRAG}.
Integrating the differential cross section over a range of $z$, as is more
practical experimentally, the resulting ratios for cuts of $z > 0.3$ and
$z > 0.5$ are shown in Fig.~4 at $Q^2 \sim 5$~GeV$^2$ for two different
asymptotic $x \rightarrow 1$ behaviors~\cite{MTNP,ISGUR_V,FJ}:
$d/u \rightarrow 0$ (dashed) and $d/u \rightarrow 1/5$ (solid).

\begin{figure}[ht]
\centering{\ \psfig{figure=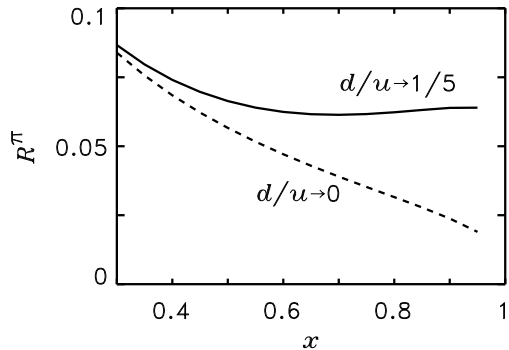,height=5.5cm}}
\caption{Semi-inclusive pion ratio $R^{\pi}$ as a function of $x$ for
        fixed $z \approx 1$.  The dashed line represents the ratio
	constructed from the CTEQ parameterization~\protect\cite{CTEQ},
	while the solid includes the modified $d$ distribution.}
\end{figure}

The HERMES Collaboration has previously extracted the $d/u$ ratio
from the $\pi^+$--$\pi^-$ difference using both proton and deuteron
targets~\cite{HERMES_DU} to increase statistics.
The advantage of using both $p$ and $d$ is that all dependence on
fragmentation functions cancels, removing any uncertainty that might
be introduced by incomplete knowledge of the hadronization process.
On the other hand, at large $x$ one still must take into account the
nuclear binding and Fermi motion effects in the deuteron, and beyond
$x \sim 0.7$ the difference between the ratios with corrected for
nuclear effects and those which are not can be quite
dramatic~\cite{SEMIPI}.
Consequently a $d/u$ ratio obtained from such a measurement without
nuclear corrections could potentially give misleading results.

On the other hand, with the high luminosity electron beam available
at CEBAF, one will be able to compensate for the falling production
rates at large $x$ and $z$, enabling good statistics to be obtained
with protons alone.
Such a measurement will become feasible with an upgraded 12 GeV
electron beam, which will enable greater access to the region of
large $Q^2$ and $W^2$.
Since the $x \rightarrow 1$ behavior is one of the very few predictions
for the $x$-dependence of quark distributions which can be drawn directly
from QCD, the results of such measurements would clearly be of enormous
interest.

\subsubsection{Quark-Hadron Duality}

Quark--hadron duality provides a beautiful illustration of the
connection between structure functions and nucleon resonance form
factors~$^{59-63}$.
In particular, it allows the behavior of valence structure functions
in the limit $x \rightarrow 1$ to be determined from the $Q^2$
dependence of elastic form factors~\cite{BG,MEL}.

The subject of quark--hadron duality, and the relation between exclusive
and inclusive processes, is actually as old as the first deep-inelastic
scattering experiments themselves.
In the early 1970s the inclusive--exclusive connection was studied
in the context of deep-inelastic scattering in the resonance region
and the onset of scaling behavior.
In their pioneering analysis, Bloom and Gilman~\cite{BG} observed that
the inclusive $F_2$ structure function at low $W$ generally follows  
a global scaling curve which describes high $W$ data, to which the
resonance structure function averages.
Furthermore, the equivalence of the averaged resonance and scaling
structure functions appears to hold for each resonance, over restricted
regions in $W$, so that the resonance---scaling duality also exists
locally~\cite{F2JL}.

Following Bloom and Gilman's empirical observations, de R\'ujula, 
Georgi and Politzer~\cite{RUJ} pointed out that global duality can
be understood from an operator product expansion of QCD moments of
structure functions.
Expanding the $F_2$ moments in a power series in $1/Q^2$
(c.f. Eq.(\ref{momdef})),
\begin{eqnarray}  
\label{htope}
\int_0^1 d\xi\ \xi^{N-2} F_2\left(\xi,Q^2\right)
&=& \sum_{k=0}^{\infty}
        \left( {(N-2) \Lambda^2 \over Q^2}
        \right)^k
A_N^{(k)} \left( \alpha_s(Q^2) \right)\ ,
\end{eqnarray}
where $\Lambda$ is some mass scale, and the Nachtmann scaling variable
$\xi = 2x / (1 + \sqrt{1 + x^2/\tau})$ takes into account target mass
corrections, one can attribute the existence of global duality to the
relative size of higher twists in deep-inelastic scattering.
The $Q^2$ dependence of the coefficients $A_N^{(k)}$ arises only through
$\alpha_s(Q^2)$ corrections, and the higher twist matrix elements
$A_N^{(k>0)}$ are expected to be of the same order of magnitude as the
leading twist term $A_N^{(0)}$.
The weak $Q^2$ dependence of the low $F_2$ moments can then be
interpreted as indicating that higher twist ($1/Q^{2k}$ suppressed)
contributions are either small or cancel.

Although global Bloom--Gilman duality of low structure function moments
can be analyzed systematically within a perturbative operator product  
expansion, an elementary understanding of local duality's origins is
more elusive.
This problem is closely related to the question of how to build up a
scaling ($\approx Q^2$ independent) structure function from resonance
contributions~\cite{IJMV}, each of which is described by a form factor 
$G_R(Q^2)$ that falls off as some power of $1/Q^2$.

To illustrate the interplay between resonances and scaling functions,
one can observe~\cite{BG,CM} that (in the narrow resonance approximation)
if the contribution of a resonance of mass $M_R$ to the $F_2$ structure   
function at large $Q^2$ is given by
(c.f. Eqs.(\ref{sfelf1})-(\ref{sfelf2}))
$F_2^{(R)} = 2 M \nu \left( G_R(Q^2) \right)^2\ \delta(W^2~-~M_R^2)$,
then a form factor behavior $G_R(Q^2) \sim (1/Q^2)^n$ translates into
a structure function
$F_2^{(R)} \sim (1-x_R)^{2n-1}$, where $x_R = Q^2/(M_R^2 - M^2 + Q^2)$.   
On purely kinematical grounds, therefore, the resonance peak at $x_R$  
does not disappear with increasing $Q^2$, but rather moves towards
$x=1$.

For elastic scattering, the connection between the $1/Q^2$ power of the
elastic form factors at large $Q^2$ and the $x \to 1$ behavior of
structure functions was first established by Drell and Yan~\cite{DY}   
and West~\cite{WEST}.
Although derived before the advent of QCD, the Drell-Yan---West
form factor--structure function relation can be expressed in perturbative
QCD language in terms of hard gluon exchange.
The pertinent observation is that deep-inelastic scattering at $x \sim 1$
probes a highly asymmetric configuration in the nucleon in which one of
the quarks goes far off-shell after exchange of at least two hard gluons
in the initial state; elastic scattering, on the other hand, requires at
least two gluons in the final state to redistribute the large $Q^2$  
absorbed by the recoiling quark~\cite{LB}.

If the inclusive--exclusive connection via local duality is taken
seriously, one can use measured structure functions in the
resonance region at large $\xi$ to directly extract elastic form
factors~\cite{RUJ}.
Conversely, empirical electromagnetic form factors at large $Q^2$ can
be used to predict the $x \to 1$ behavior of deep-inelastic structure
functions~\cite{BG}.

Integrating the elastic contributions to the structure functions in
Eqs.(\ref{sfelf1})-(\ref{sfelg2}) over the Nachtmann variable $\xi$,
where $\xi = 2 x / (1 + \sqrt{1 + x^2/\tau})$, between the pion
threshold $\xi_{th}$ and $\xi=1$, one finds `localized' moments
of the structure functions:
\begin{eqnarray}
\label{NmomF1}
\int_{\xi_{th}}^1  d\xi\ \xi^{N-2}\ F_1(\xi,Q^2)
&=& { \xi_0^N \over 4 - 2 \xi_0 }\ G_M^2(Q^2)\ ,       \\
\label{NmomF2}
\int_{\xi_{th}}^1  d\xi\ \xi^{N-2}\ F_2(\xi,Q^2)
&=& { \xi_0^N \over 2 - \xi_0 }\
    { G_E^2(Q^2) + \tau G_M^2(Q^2) \over 1 + \tau }\ ,     \\
\label{NmomG1}
\int_{\xi_{th}}^1  d\xi\ \xi^{N-2}\ g_1(\xi,Q^2)
&=& { \xi_0^N \over 4 - 2 \xi_0 }\
    { G_M(Q^2) \left( G_E(Q^2) + \tau G_M(Q^2) \right)
      \over 1 + \tau }\ ,                                  \\
\label{NmomG2}  
\int_{\xi_{th}}^1  d\xi\ \xi^{N-2}\ g_2(\xi,Q^2)
&=& { \xi_0^N \over 4 - 2 \xi_0 }\
    { \tau G_M(Q^2) \left( G_E(Q^2) - G_M(Q^2) \right)
    \over 1 + \tau }\ ,
\end{eqnarray}
where $\tau = Q^2/4M^2$ and $\xi_0 = 2 / (1 + \sqrt{1 + 1/\tau})$ is
the value of $\xi$ at $x=1$.
Differentiating Eqs.(\ref{NmomF1})--(\ref{NmomG2}) with respect to $Q^2$
for $N=2$ allows the inclusive structure functions near $x=1$ to be  
extracted from the elastic form factors and their
$Q^2$-derivatives~\cite{MEL}:
\begin{eqnarray}
\label{SFdualf1}
F_1 &\propto&
{ dG_M^2 \over dQ^2 }\ ,                                \\
\label{SFdualf2}
F_2
&\propto& { G_M^2 - G_E^2 \over 4 M^2 (1+\tau)^2 }
+ { 1 \over 1 + \tau }
  \left( { dG_E^2 \over dQ^2 } + \tau { dG_M^2 \over dQ^2 }
  \right)\ ,                                             \\
\label{SFdualg1}
g_1
&\propto& { G_M \left( G_M - G_E \right) \over 4 M^2 (1+\tau)^2 }
+ { 1 \over 1 + \tau }
  \left( { d(G_E G_M) \over dQ^2 } + \tau { dG_M^2 \over dQ^2 }
  \right)\ ,                                             \\
\label{SFdualg2}
g_2
&\propto& { G_M \left( G_M - G_E \right) \over 4 M^2 (1+\tau)^2 }
+ { \tau \over 1 + \tau }
  \left( { d(G_E G_M) \over dQ^2 } + { dG_M^2 \over dQ^2 }
  \right)\ .
\end{eqnarray}
Note that as $\tau \to \infty$ each of the structure functions $F_1$,
$F_2$ and $g_1$ is determined by the slope of the square of the magnetic
form factor, while $g_2$ (which in deep-inelastic scattering is associated
with higher twists) is determined by a combination of $G_E$ and $G_M$.

Equations (\ref{SFdualf1})-(\ref{SFdualg2}) allow the $x \sim 1$
behavior of structure functions to be predicted from empirical
electromagnetic form factors.
The ratios of the neutron to proton $F_1$, $F_2$ and $g_1$ structure
functions are shown in Fig.~5 as a function of $Q^2$, using typical
parameterizations~\cite{MDM} of the global form factor data.
While the $F_2$ ratio varies quite rapidly at low $Q^2$, beyond
$Q^2 \sim 3$~GeV$^2$ it remains almost $Q^2$ independent, approaching
the asymptotic value $(dG_M^{n 2}/dQ^2)/(dG_M^{p 2}/dQ^2)$.
This is consistent with the operator product expansion interpretation
of de R\'ujula et al.~\cite{RUJ} in which duality should be a better
approximation with increasing $Q^2$.
Because the $F_1^n/F_1^p$ ratio depends only on $G_M$, it remains
flat over nearly the entire range of $Q^2$.
At asymptotic $Q^2$ the model predictions for $F_1(x\to 1)$ coincide
with those for $F_2$; at finite $Q^2$ the difference between $F_1$ and  
$F_2$ can be used to predict the $x \to 1$ behavior of the longitudinal
structure function, or the $R=\sigma_L/\sigma_T$ ratio.

The pattern of SU(6) breaking for the spin-dependent structure function
ratio $g_1^n/g_1^p$ essentially follows that for $F_2^n/F_2^p$, namely
1/4 in the $d$ quark suppression and 3/7 in the helicity flip
suppression scenarios~\cite{MTNP,ISGUR_V}.
However, the $g_1$ structure function ratio approaches the asymptotic  
limit somewhat more slowly than $F_1$ or $F_2$, which may indicate a
more important role played by higher twists in spin-dependent structure
functions than in spin-averaged.

It appears to be an interesting coincidence that the helicity retention
model prediction of 3/7 is very close to the empirical ratio of the
squares of the neutron and proton magnetic form factors,
$\mu_n^2/\mu_p^2 \approx 4/9$.
Indeed, if one approximates the $Q^2$ dependence of the proton and
neutron form factors by dipoles, and takes $G_E^n \approx 0$, then the
structure function ratios are all given by simple analytic expressions,
$F_2^n/F_2^p \approx F_1^n/F_1^p \approx g_1^n/g_1^p \to \mu_n^2/\mu_p^2$
as $Q^2 \to \infty$.
On the other hand, for the $g_2$ structure function, which depends on
both $G_E$ and $G_M$ at large $Q^2$, one has a different asymptotic
behavior, $g_2^n/g_2^p \to \mu_n^2 / (\mu_p (1 + \mu_p)) \approx 0.345$.

\begin{figure}[ht]
\centering{\ \psfig{figure=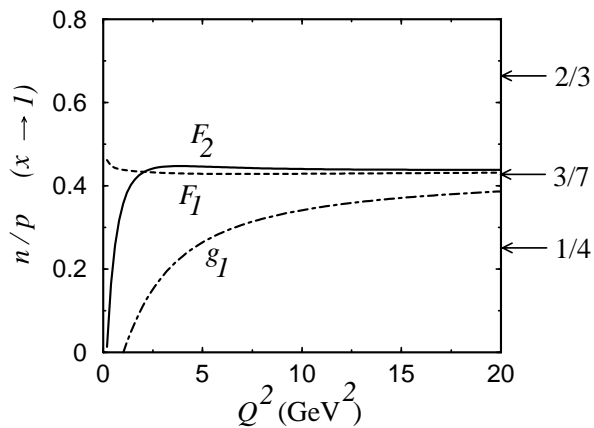,height=5.5cm}}
\caption{Neutron to proton ratio for $F_1$ (dashed), $F_2$ (solid)
	and $g_1$ (dot-dashed) structure functions in the limit
	$x \rightarrow 1$.}
\end{figure}

If the resonance structure functions at large $\xi$ are known,
one can conversely extract the nucleon electromagnetic form factors
from Eqs.(\ref{NmomF1})--(\ref{NmomG2}).  
The $G_M$ form factor of the nucleon can be extracted directly from
the measured $F_1(\xi,Q^2)$ structure function in Eq.(\ref{NmomF1}).
Unfortunately, only the $F_2(\xi,Q^2)$ structure function of the proton
has so far been measured in the resonance region.
Nevertheless, to a good approximation one can assume that the ratio of
electric to magnetic form factors is reasonably well known (see, however,
Ref.~\cite{GEMJL}), and extract $G_M$ from the $F_2$ structure function
in the resonance region via Eq.(\ref{NmomF2}).

Using the parameterization of the recent $F_2(\xi,Q^2)$ data from
Jefferson Lab~\cite{F2JL}, in Fig.~6 we show the extracted $G_M^p$
compared with a compilation of elastic data.
The agreement with data is quite remarkable over the entire range
of $Q^2$ between 0 and 3 GeV$^2$.

\begin{figure}  
\centering{\ \psfig{figure=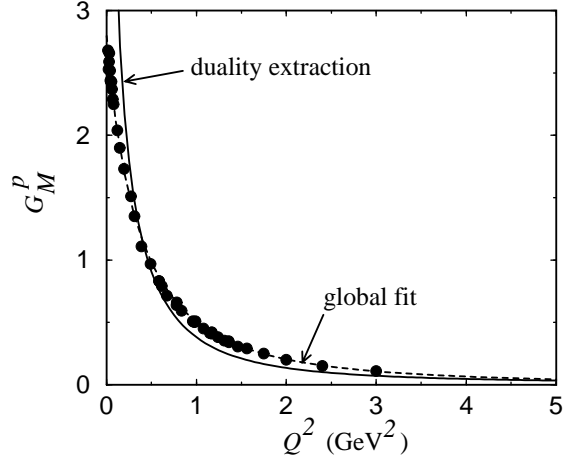,height=6cm}}
\caption{Proton magnetic form factor extracted from the inclusive
        structure function via Eq.(\protect\ref{NmomF2}).}
\end{figure}

The reliability of the duality predictions is of course only as good as
the quality of the empirical data on the electromagnetic form factors
and resonance structure functions.
While the duality relations~\cite{RUJ} are expected to be progressively
more accurate with increasing $Q^2$, the difficulty in measuring form
factors at large $Q^2$ also increases.
Experimentally, the proton magnetic form factor $G_M^p$ is relatively
well constrained to $Q^2 \sim 30$ GeV$^2$, and the proton electric
$G_E^p$ to $Q^2 \sim 10$ GeV$^2$.
The neutron magnetic form factor $G_M^n$ has been measured to
$Q^2 \sim 5$~GeV$^2$, although the neutron $G_E^n$ is not very well
determined at large $Q^2$ (fortunately, however, this plays only a
minor role in the duality relations, with the exception of the neutron
to proton $g_2$ ratio, Eq.(\ref{SFdualg2})).

Obviously more data at larger $Q^2$ would allow more accurate
predictions for the $x \to 1$ structure functions, and new experiments
at Jefferson Lab~\cite{GEMJL} will provide valuable constraints.
Once data on the longitudinal and spin-dependent structure functions at
large $x$ become available, a more complete test of local duality between
elastic form factors and $x \sim 1$ structure functions can be made.

\subsection{Light Quark Sea}
 
Over the past decade a number of high-energy experiments and refined
data analyses have forced a re-evaluation of our view of the nucleon
in terms of three valence quarks immersed in a sea of perturbatively
generated $q\bar q$ pairs and gluons~\cite{SCIENCE}.
A classic example of this is the asymmetry of the light quark sea of
the proton, dramatically confirmed in recent deep-inelastic and Drell-Yan
experiments at CERN~\cite{NMCGSR,NA51} and Fermilab~\cite{E866}.

Difference between quark or antiquark distributions in the proton sea
almost universally signal the presence of phenomena which require
understanding of strongly coupled QCD.
Their existence testifies to the relevance of long-distance dynamics
(which are responsible for confinement) even at large energy and
momentum transfers.

Because gluons in QCD are flavor-blind, the perturbative process
$g \rightarrow q \bar q$ gives rise to a sea component of the
nucleon which is symmetric in the quark flavors.
Although differences can arise due to different quark masses, because
isospin symmetry is such a good symmetry in nature, one would expect
that the sea of light quarks generated perturbatively would be almost
identical, $\bar u(x) = \bar d(x)$.

It was therefore a surprise to many when measurements by the New Muon
Collaboration (NMC) at CERN~\cite{NMCGSR} of the proton and deuteron
structure functions suggested a significant excess of $\bar d$ over
$\bar u$ in the proton.
Indeed, it heralded a renewed interest in the application of ideas
from non-perturbative QCD to deep-inelastic scattering analyses.
While the NMC experiment measured the integral of the antiquark
difference, more recently the E866 experiment at Fermilab has for
the first time mapped out the shape of the $\bar d / \bar u$ ratio
over a large range of $x$, $0.02 < x < 0.345$.

Specifically, the E866/NuSea Collaboration measured $\mu^+\mu^-$
Drell-Yan pairs produced in $pp$ and $pd$ collisions.
In the parton model the Drell-Yan cross section is proportional to:
\begin{eqnarray}
\sigma^{ph}
&\propto& \sum_q e_q^2
\left( q^p(x_1)\ \bar q^h(x_2) + \bar q^p(x_1)\ q^h(x_2) \right),
\end{eqnarray}  
where $h = p$ or $D$, and $x_1$ and $x_2$ are the light-cone momentum
fractions carried by partons in the projectile and target hadron,
respectively.
Using isospin symmetry to relate quark distributions in the neutron
to those in the proton, in the limit $x_1 \gg x_2$ (in which
$\bar q(x_1) \ll q(x_1)$) the ratio of the deuteron to proton cross
sections can be written:
\begin{eqnarray}
\label{nonuc}
{ \sigma^{pD} \over 2 \sigma^{pp} }
&=& {1 \over 2}
\left( 1 + { \bar d(x_2) \over \bar u(x_2) } \right)
{ 4 + d(x_1)/u(x_1) \over
  4 + d(x_1)/u(x_1) \cdot \bar d(x_2)/\bar u(x_2) }\ ,
\end{eqnarray}
Corrections for nuclear shadowing in the deuteron~\cite{MTSHAD,TM},
which are important at $x \ll 0.1$, are small in the region covered
by this experiment.

The relatively large asymmetry found in these experiments, shown in
Fig.~7 implies the presence of non-trivial dynamics in the proton
sea which does not have a perturbative QCD origin.
The simplest and most obvious source of a non-perturbative asymmetry
in the light quark sea is the chiral structure of QCD.
{}From numerous studies in low energy physics, including chiral
perturbation theory~\cite{CHIPT}, pions are known to play a crucial
role in the structure and dynamics of the nucleon (see Section~2).
However, there is no reason why the long-range tail of the nucleon
should not also play a role at higher energies.

\begin{figure}[ht]
\centering{\ \psfig{figure=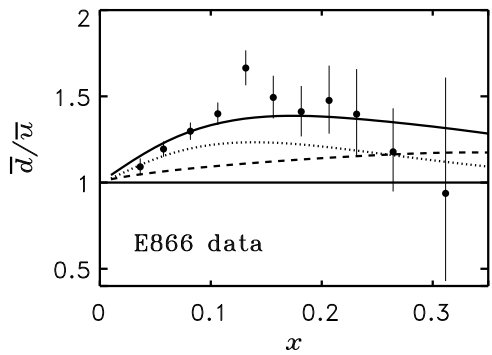,height=5.5cm}}
\caption{Flavor asymmetry of the light antiquark sea, including pion
	cloud (dashed) and Pauli blocking effects (dotted), and the
	total (solid).}
\end{figure}

As pointed out by Thomas \cite{AWT83}, if the proton's wave function
contains an explicit $\pi^+ n$ Fock state component, a deep-inelastic
probe scattering from the virtual $\pi^+$, which contains a valence
$\bar d$ quark, will automatically lead to a $\bar d$ excess in the
proton.
To be specific, consider a model in which the nucleon core consists of
valence quarks, interacting via gluon exchange for example, with sea
quark effects introduced through the coupling of the core to $q \bar q$
states with pseudoscalar meson quantum numbers (many variants of such a
model exist --- see for example Refs.\cite{CBM,GI}).
The physical nucleon state (momentum $P$) can then be expanded
(in the one-meson approximation) as a series involving bare nucleon
and two-particle meson--baryon states:
\begin{eqnarray}
\label{Fock}
\left| N(P) \rangle_{\rm phys} \right.
&=& \sqrt{Z}\
\left\{ \left| N(P) \rangle_{\rm bare} \right. \right.       \nonumber\\   
& & \hspace*{-2.5cm} +\  \sum_{B,M} \int dy\ d^2{\bf k}_\perp\
     g_{MNB}\ \phi_{BM}(y,{\bf k}_\perp)\
     \left| B (y,{\bf k}_\perp); M (1-y,-{\bf k}_\perp) \rangle \right.
\left.
\right\}\ ,
\end{eqnarray}
where $M=\pi,K,\cdots$ and $B=N,\Delta,\Lambda,\cdots$.
The function $\phi_{BM}(y,{\bf k}_\perp)$ is the probability amplitude
for the physical nucleon $N$ to be in a state consisting of a baryon
$B$ and meson $M$, having transverse momenta ${\bf k}_\perp$ and
$-{\bf k}_\perp$, and carrying longitudinal momentum fractions
$y = k_+/P_+$ and 1--$y = (P_+ - k_+)/P_+$, respectively.
The bare nucleon probability is denoted by $Z$, and $g_{MNB}$ is the
$MNB$ coupling constant.
The one-meson approximation in Eq.(\ref{Fock}) will be valid as long
as the meson cloud is relatively soft ($Z \approx 1$).
It will progressively break down for harder $MNB$ vertex functions,
at which point one will need to include two-meson and higher order
Fock state components in Eq.(\ref{Fock}).

Because of their small masses, the $\pi N$ component of the Fock state
expansion in Eq.(\ref{Fock}) will be the most important (contributions
from heavier mesons and baryons will be progressively suppressed with
increasing mass).
In the impulse approximation, deep-inelastic scattering from the $\pi N$
component of the proton can then be understood in the IMF as the
probability for a pion to be emitted by the proton, folded with the
probability of finding the a parton in the pion~\cite{IMF}.
For the antiquark asymmetry, this can be written as~\cite{REV}:
\begin{eqnarray}
\label{conv}
\bar d(x) - \bar u(x)
&=& {2 \over 3} \int_x^1 {dy \over y} f_{\pi N}(y)\ \bar q^\pi (x/y)\ ,
\end{eqnarray}
where $\bar q^\pi$ is the (valence) quark distribution in the pion
(e.g. $\bar u$ in $\pi^+$, normalized to unity), and the distribution
of pions with a recoiling nucleon ($N \rightarrow \pi n$ splitting
function) is given by~$^{75,80-82}$:
\begin{eqnarray}
\label{fypin}
f_{\pi N}(y)
&=&
{ 3 g^2_{\pi N N} \over 16 \pi^3 }
\int { d^2{\bf k}_T \over (1-y) }
\frac{ {\cal F}_{\pi N}^2(s_{\pi N}) }
     { y\ (M^2 - s_{\pi N})^2 }
\left( { k_T^2 + y^2 M^2 \over 1-y } \right)\ .
\end{eqnarray}
For the $\pi NN$ vertex a pseudoscalar $i\gamma_5$ interaction has
been used, although in the IMF the same results are obtained with a
pseudovector coupling.
The invariant mass squared of the $\pi N$ system is given by:
\begin{eqnarray}
\label{s_piN}
s_{\pi N} &=& {k^2_T + m_{\pi}^2 \over y}
	   +  {k^2_T + M^2 \over 1-y}\ ,
\end{eqnarray}
and for the functional form of the $\pi NN$ vertex function
${\cal F}_{\pi N}(s_{\pi N})$ we can take a simple dipole
parameterization,
${\cal F}_{\pi N}(s_{\pi N})
 = \left( ( \Lambda^2 + M^2 )
	/ ( \Lambda^2 + s_{\pi N} )
   \right)^2$,
normalized so that the coupling constant $g_{\pi N N}$ has its  
standard value (= 13.07) at the pole (${\cal F}(M^2)~=~1$).
Note that the contribution from the pion cloud in Eq.(\ref{conv})
is a leading twist effect, which scales with $Q^2$
(at leading order the $Q^2$ dependence of $\bar d-\bar u$ in
Eq.(\ref{conv}) enters through the leading-twist quark distribution
in the pion, $\bar q^\pi$.

One can easily generalize the above to include higher Fock state
components~\cite{REV}, most notably the $\Delta$.
Because the dominant process there is $p \rightarrow \Delta^{++} \pi^-$,
the $\Delta$ will actually cancel some of the $\bar d$ excess generated
through the $\pi N$ component, although this will be somewhat smaller
due to the larger mass of the $\Delta$.

The relative contributions are partly determined by the $\pi NN$ and
$\pi N\Delta$ vertex form factor.
The form factor cut-offs $\Lambda$ can be determined phenomenologically
by comparing against various inclusive and semi-inclusive
data~\cite{HEAVY}, although the most direct way to fix these parameters
is through a comparison of the axial form factors for the nucleon and
for the $N$--$\Delta$ transition.
Within the framework of PCAC these form factors are directly related
to the corresponding form factors for pion emission or absorption.
The data on the axial form factor are best fit, in a dipole
parameterization, by a 1.3~(1.02)~GeV dipole for the axial $N$
($N$--$\Delta$ transition) form factor~\cite{AXIAL}, which gives
a pion probability in the proton of $\approx 13\% (10\%)$.

With these parameters Fig~.6 shows the $\bar d/\bar u$ ratio in the
proton due to $\pi N$ and $\pi \Delta$ components of the nucleon
wave function (dashed line)~\cite{DYN}.
Data~\cite{CTEQ} on the sum of the $\bar u$ and $\bar d$ (which is
dominated by perturbative contributions) has been used to convert
the calculated $\bar d-\bar u$ difference to the $\bar d/\bar u$ ratio.
The results suggest that with pions alone one can account for about half
of the observed asymmetry, leaving room for possible contributions from
other mechanisms.

Another mechanism which could also contribute to the $\bar d-\bar u$
asymmetry is associated with the effects of antisymmetrization of
$q \bar q$ pairs created inside the core~\cite{SST,FMST}.
As pointed out originally by Field and Feynman~\cite{FF}, because the
valence quark flavors are unequally represented in the proton, the Pauli
exclusion principle will affect the likelihood with which $q\bar q$
pairs can be created in different flavor channels.
Since the proton contains 2 valence $u$ quarks compared with only one
valence $d$ quark, $u \bar u$ pair creation will be suppressed relative
to $d \bar d$ creation.
In the ground state of the proton the suppression will be in the
ratio $\bar d : \bar u = 5:4$.

Phenomenological analyses in terms of low energy models (specifically,
the MIT bag model~\cite{BAG}) suggest that the contribution from Pauli
blocking can be parameterized as
$(\bar d~-~\bar u)^{\rm Pauli}
= \tau^{\rm Pauli} (\alpha+1) (1-x)^\alpha$,
where $\alpha$ is some large power, with normalization,
$\tau^{\rm Pauli}$, less than $\approx 25\%$.
Phenomenologically, one finds a good fit with $\alpha \approx 14$
and a normalization $\tau^{\rm Pauli} \approx 7\%$, which is at the
lower end of the expected scale but consistent with the bag model
predictions~\cite{BAG}.
Together with the integrated asymmetry from pions,
$\tau^{\pi} \sim 0.05$, the combined value
$\tau = \tau^{\pi} + \tau^{\rm Pauli} \approx 0.12$ is in quite
reasonable agreement with the experimental result, $0.100 \pm 0.018$
from E866.

Although the combined pion cloud and Pauli blocking mechanisms are able
to fit the E866 data reasonable well at small and intermediate $x$
($x < 0.2$), it is difficult to reproduce the apparent trend in the
data at large $x$ towards zero asymmetry, and possibly even an excess
of $\bar u$ for $x > 0.3$.
Unfortunately, the error bars are quite large beyond $x \sim 0.25$,
and it is not clear whether any new Drell-Yan data will be forthcoming
in the near future to clarify this.

A solution might be available, however, through semi-inclusive scattering,
tagging charged pions produced off protons and neutrons.
Taking the ratio of the isovector combination of cross sections for
$\pi^+$ and $\pi^-$ production~\cite{LMS}:
\begin{eqnarray}
{ N_p^{\pi^+ + \pi^-} - N_n^{\pi^+ + \pi^-} \over
  N_p^{\pi^+ - \pi^-} - N_n^{\pi^+ - \pi^-} }
&=& {3 \over 5}
    \left( { u - d - \bar d + \bar u  \over  u - d  + \bar d - \bar u}
    \right)
    \left( { D + \overline D \over D - \overline D } \right)\ .
\end{eqnarray}
the difference $\bar d-\bar u$ can be directly measured provided
the $u$ and $d$ quark distributions and fragmentation functions
are known.
The HERMES Collaboration has in fact recently measured this
ratio~\cite{HERMES_DUBAR}, although there the rapidly falling
cross sections at large $x$ make measurements beyond $x \sim 0.3$
challenging.
On the other hand, a high luminosity electron beam such as that
available at Jefferson Lab, could, with higher energy, allow the
asymmetry to be measured well beyond $x \sim 0.3$ with relatively
small errors.
This would parallel the semi-inclusive measurement of the $d/u$ ratio
through Eq.(\ref{Rpi}) at somewhat larger~$x$.

\begin{figure}[h]
\centering{\ \psfig{figure=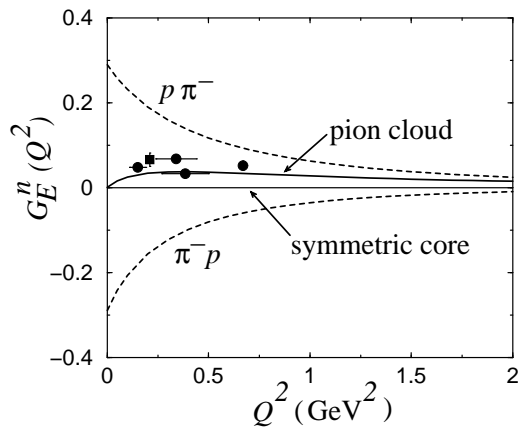,height=5.5cm}}
\caption{Neutron electric form factor in the pion cloud
	model~\protect\cite{EXCL}.
	The direct $\pi^-$ coupling contribution is labeled
	``$\pi^- p$'', and the recoil proton ``$p \pi^-$''.}
\end{figure}

If a pseudoscalar cloud of $q \bar q$ states plays an important role
in the $\bar d/\bar u$ asymmetry, its effects should also be visible
in other flavor-sensitive observables, such as electromagnetic form
factors~\cite{EXCL}.
An excellent example is the electric form factor of the
neutron~\cite{FFN}, a non-zero value for which can arise from
a pion cloud, $n \rightarrow p \pi^-$.
Although in practice other effects~$^{92-94}$ such as spin-dependent
interactions due to one gluon exchange between quarks in the core will
certainly contribute at some level, it is important nevertheless to
test the consistency of the above model by evaluating its consequences
for all observables that may carry its signature.

To illustrate the sole effect of the pion cloud, all residual interactions
between quarks in the core can be switched off, so that the form factors
have only two contributions: one in which the photon couples to the
virtual $\pi$ and one where the photon couples to the recoil nucleon:
\begin{eqnarray}
{\cal F}_{1,2}(Q^2)  
&=& \int_0^1 dy\
\left( f_{1,2}^{(\pi)}(y,Q^2)\
    +\ f_{1,2}^{(N)}(y,Q^2)
\right)\ .
\end{eqnarray}
The recoil nucleon contribution is described by the functions:
\begin{eqnarray}  
\label{f1N}
f_1^{(N)}(y,Q^2)
&=&
{ 3 g_{\pi NN}^2 \over 16 \pi^3 }
\int { d^2{\bf k}_\perp \over y^2 (1-y) }
{ {\cal F}(s_{N\pi,i})\ {\cal F}(s_{N\pi,f})
  \over (s_{N\pi,i} - M^2) (s_{N\pi,f} - M^2) }		\nonumber\\
& & \times
\left(
  k_\perp^2 + M^2 (1 - y)^2 - (1-y)^2 { q_\perp^2 \over 4 }
\right)\ ,                                        	\\
\label{f2N}
f_2^{(N)}(y,Q^2)
&=&
{ 3 g_{\pi NN}^2 \over 16 \pi^3 }
\int { d^2{\bf k}_\perp \over y^2 (1-y) }
{ {\cal F}(s_{N\pi,i})\ {\cal F}(s_{N\pi,f})
  \over (s_{N\pi,i} - M^2) (s_{N\pi,f} - M^2) }		\nonumber\\
& & \times (-2 M^2) (1-y)^2\ ,
\end{eqnarray}
where the squared center of mass energies are:
\begin{eqnarray}
s_{N\pi,i(f)}
&=& s_{\pi N}\
 +\ {{\bf q}_\perp \over y} \cdot
    \left( (1-y) {{\bf q}_\perp \over 4} \pm {\bf k}_\perp \right)\ ,
\end{eqnarray}
with $s_{\pi N}$ defined in Eq.(\ref{s_piN}).

The contribution from coupling directly to the pion is:
\begin{eqnarray}
\label{f1pi}
f_1^{(\pi)}(y,Q^2)  
&=&
{ 3 g_{\pi NN}^2 \over 16 \pi^3 }
\int { d^2{\bf k}_\perp \over y^2 (1-y) }
{ {\cal F}(s_{\pi N,i})\ {\cal F}(s_{\pi N,f})
  \over (s_{\pi N,i} - M^2) (s_{\pi N,f} - M^2) }	\nonumber \\
& & \times
\left( k_\perp^2 + M^2 (1-y)^2
     - y^2 {q_\perp^2 \over 4}
\right)\ ,						\\
\label{f2pi}
f_2^{(\pi)}(y,Q^2)
&=&
{ 3 g_{\pi NN}^2 \over 16 \pi^3 }
\int { d^2{\bf k}_\perp \over y^2 (1-y) }
{ {\cal F}(s_{\pi N,i})\ {\cal F}(s_{\pi N,f})
  \over (s_{\pi N,i} - M^2) (s_{\pi N,f} - M^2) }	\nonumber \\
& & \times
\left( 2 M^2 y (1-y) \right)\ ,
\end{eqnarray}
where the $\pi N$ squared center of mass energies are:
\begin{eqnarray}
s_{\pi N,i(f)}
&=& s_{\pi N}\
 +\ {{\bf q}_\perp \over 1-y} \cdot
    \left( y\ {{\bf q}_\perp \over 4} \pm {\bf k}_\perp \right)\ .
\end{eqnarray}
The $N \rightarrow \pi N$ splitting functions are related to the
distribution functions in Eqs.(\ref{f1N}) and (\ref{f1pi}) by:
\begin{eqnarray}
f_1^{(\pi)}(y,Q^2=0) &=& f_{\pi N}(y)\ ,	\\
f_1^{(N)}(y,Q^2=0)   &=& f_{N \pi}(y)\ =\ f_{\pi N}(1-y)\ .
\end{eqnarray}
Using the same pion cloud parameters as in the calculation of the
$\bar d/\bar u$ asymmetry in Fig.~7, the relative contributions
to $G_E^n$ from the $\pi^-$ and recoil proton are shown in Fig.~8.
Both are large in magnitude but opposite in sign, so that the combined
effects cancel to give a small positive $G_E^n$, consistent with
the data.
Note, however, that the Pauli blocking effect plays no role in
form factors, since any suppression of $\bar u$ relative to $\bar d$
here would be accompanied by an equal and opposite suppression of
$u_{\rm sea}$ relative to $d_{\rm sea}$, and form factors always
contain charge conjugation odd (valence) combinations of flavors.

The fact that the model prediction underestimates the strength of the
observed $G_E^n$ suggests that other mechanisms, such as the color
hyperfine interaction generated by one-gluon exchange between quarks in
the core~\cite{IKK,IKS}, are likely to be responsible for some of the
difference.
The lowest order Hamiltonian for the color-magnetic hyperfine
interaction~\cite{IKK} between two quarks is proportional to
$(\alpha_s/m_i m_j) \vec S_i \cdot \vec S_j$.
Because this force is repulsive if the spins of the quarks are parallel
and attractive if they're antiparallel, from the SU(6) wave function in
Eq.(\ref{pwfn}) it naturally leads to an increase in the mass of the
$\Delta$ and a lowering of the mass of the nucleon.
The same force also leads to the softening~\cite{ISGUR_V} of the $d$
quark distribution relative to the $u$ (see Eq.(\ref{S0dom})).
Furthermore, it leads to a distortion of the spatial (and hence charge)
distributions of quarks in the neutron, pushing the two (negatively
charged) $d$ to the periphery of the neutron, while forcing the
(positively charged) $u$ in the center, giving rise to a negative
charge radius~\cite{IKS}, $\left< \sum_i e_i r_i^2 \right>_n$.

\begin{figure}[h]
\centering{\ \psfig{figure=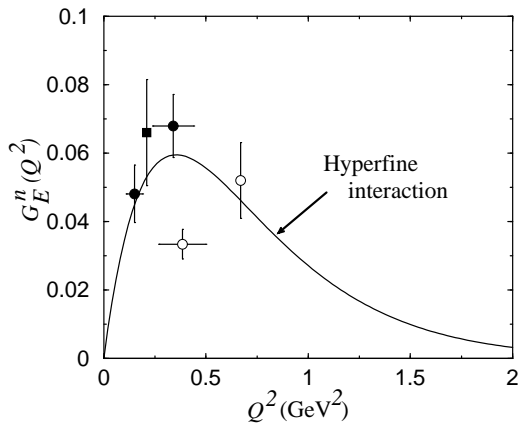,height=5.5cm}}
\caption{Electric neutron form factor in the hyperfine perturbed
	quark model with a harmonic oscillator potential.}
\end{figure}

In the harmonic oscillator model at leading order in $\alpha_s$,
the hyperfine interaction gives rise to a neutron electric form
factor~\cite{IKS}:
\begin{eqnarray}
G_E^n(Q^2)
&=& -{1 \over 6} \left< \sum_i e_i r_i^2 \right>_n
	Q^2 \exp \left( -Q^2/6\alpha^2 \right)\ ,
\end{eqnarray}
where $\alpha$ can be related to decay amplitudes and charge
radii~\cite{IKK}.
Taking the value $\alpha=0.243$ from the ratio of neutron to proton
charge radii, the resulting form factor in Fig.~9 agress quite well
with the available $G_E^n$ data.
More accurate data, which will soon be available from Jefferson Lab
and elsewhere over a range of $Q^2$ will allow more systematic
comparison of the various mechanisms which contribute to SU(6)
symmetry breaking.

\subsection{Strange Quarks in the Nucleon}

A complication in studying the light quark sea is the fact that
non-perturbative features associated with $u$ and $d$ quarks are   
intrinsically correlated with the valence core of the proton,
so that effects of $q \bar q$ pairs can be difficult to distinguish 
from those of antisymmetrization or residual interactions of quarks
in the core.
The strange sector, on the other hand, where antisymmetrization
between sea and valence quarks plays no role, is therefore more
likely to provide direct information about the non-perturbative
origin of the nucleon sea~\cite{JT}.

Evidence for non-perturbative strangeness is currently being sought
in a number of processes, ranging from semi-inclusive neutrino induced
deep-inelastic scattering to parity violating electron--proton scattering.
As for the $\bar d-\bar u$ asymmetry, perturbative QCD alone generates
identical $s$ and $\overline s$ distributions, so that any asymmetry
would have to be non-perturbative in origin.

\begin{figure}[ht]
\centering{\ \psfig{figure=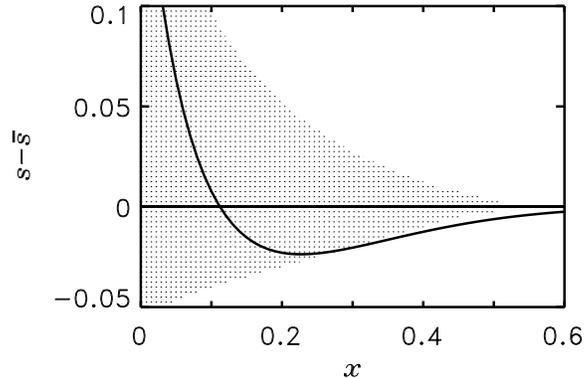,height=6cm}}
\caption{Strange quark asymmetry in the proton arising from a
	kaon cloud of the nucleon.
	The shaded region indicates current experimental limits 
	from the CCFR Collaboration~\protect\cite{CCFR}.}
\end{figure}

In deep-inelastic scattering, the CCFR collaboration~\cite{CCFR} analyzed
charm production cross sections in $\nu$ and $\bar \nu$ reactions, which
probe the $s$ and $\bar s$ distributions in the nucleon, respectively.
The resulting difference $s-\bar s$, indicated in Fig.~10 by the shaded
area, has been extracted from the $s/\bar s$ ratio and absolute values of
$s+\bar s$ from global data parameterizations.

The curve in Fig.~10 corresponds to the chiral cloud model prediction
for the asymmetry (in analogy with the pion cloud in Section~4.2), in
which the strangeness in the nucleon is carried by kaons and hyperons,
so that the $s$ and $\bar s$ quarks have quite different
origins~\cite{GI,ST}.
Taking the $\Lambda$ hyperon as an illustration (the results generalize
straightforwardly to other hyperons such as the $\Sigma$), the difference
between the $s$ and $\bar s$ can be written~\cite{MM}:
\begin{eqnarray}
s(x) - \bar s(x)
&=& \int_x^1 { dy \over y }
\left( f_{\Lambda K}(y)\ s^{\Lambda}(x/y)\
     - f_{K \Lambda}(y)\ \bar s^K(x/y)
\right)\ ,
\end{eqnarray}
where the $K$ distribution function $f_{K\Lambda}$ is the analog of the
$\pi N$ splitting function in Eq.(\ref{fypin}), and the corresponding
$\Lambda$ distribution $f_{\Lambda K}(y) = f_{K\Lambda}(1-y)$.

In the IMF parameterization of the $KNY$ ($Y=\Lambda, \Sigma$) vertex
function, because the $\bar s$ distribution in a kaon is much harder than
the $s$ distribution in a hyperon, the resulting $s-\bar s$ difference
is negative at large $x$, despite the kaon distribution in the nucleon
being slightly softer than the hyperon distribution~\cite{MM}.
With a dipole cut-off mass of $\Lambda \sim 1$~GeV, the kaon probability
in the nucleon is $\approx 3\%$.
On the other hand, the exact shape and even sign of the $s-\bar s$
difference as a function of $x$ is quite sensitive to the shape of
the $KNY$ vertex~\cite{MM}.

Overall, while the current experimental $s-\bar s$ difference is
consistent with zero, it does also consistent with a small amount
of non-perturbative strangeness, which would be generated from a
kaon cloud around the nucleon~\cite{MM}.
Of course other, heavier strange mesons and hyperons can be added to the
analysis, although in the context of chiral symmetry~\cite{LNA} the
justification for inclusion of heavier pseudoscalar as well as vector
mesons is less clear.
The addition of the towers of heavier mesons and baryons has also been
shown in a quark model~\cite{GI} to lead to significant cancellations,
leaving the net strangeness in the nucleon quite small.

\begin{figure}[h]
\centering{\ \psfig{figure=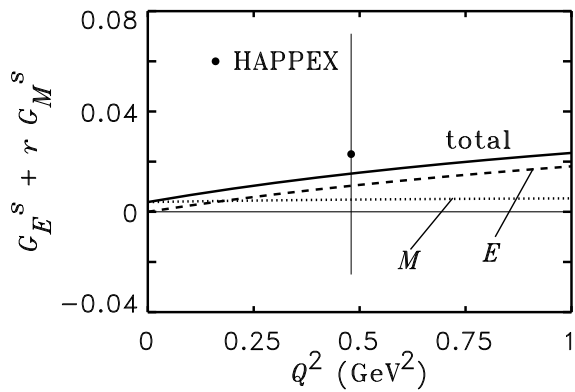,height=6cm}}
\caption{Strange electromagnetic form factors of the proton compared
	with a kaon cloud prediction, with the magnetic ($M$),
	electric ($E$) and total contributions indicated.
	For the HAPPEX data~\protect\cite{HAPPEX}, $r\approx 0.4$.}
\end{figure}

Within the same formalism as used to discuss strange quark distributions
one can also calculate the strangeness form factors of the nucleon,
which are being measured in parity-violating electron scattering
experiments at MIT-Bates~\cite{SAMPLE} and Jefferson Lab~\cite{HAPPEX}.
The HAPPEX Collaboration at Jefferson Lab~\cite{HAPPEX} has recently
measured the left-right asymmetry $A_{LR}$ in $\vec e p \rightarrow e p$
elastic scattering, which measures the $\gamma^* Z$ interference term:
\begin{eqnarray}
A &=& { \sigma_R - \sigma_L \over \sigma_R + \sigma_L }
\  =\
\left( { - G_F \over \pi \alpha_{em} \sqrt{2} } \right)
{ 1 \over \varepsilon\ G_E^{p 2} + \tau\ G_M^{p 2} }	\nonumber\\
  & & \hspace*{-0.5cm} \times
\left( \varepsilon\ G_E^p\ G_E^{p (Z)}
     + \tau\     G_M^p\ G_M^{p (Z)}
     - {1\over 2} (1 - 4 \sin^2\theta_W)\ \varepsilon'\ G_M^p\ G_A^{p (Z)}
\right),
\end{eqnarray}
with
$\varepsilon = \left( 1 + 2 (1+\tau) \tan^2(\theta/2) \right)^{-1}$,  
and $\varepsilon' = \sqrt{ \tau (1+\tau) (1-\varepsilon^2) }$
(the $Q^2$ dependence in all form factors is implicit).
Using isospin symmetry, one can relate the electric and magnetic form
factors for photon and $Z$-boson exchange via:
\begin{eqnarray}
G_{E,M}^{p (Z)}
&=& {1 \over 4} G_{E,M}^{(I=1)}
 -  \sin^2\theta_W\ G_{E,M}^p
 -  {1 \over 4} G_{E,M}^s\ ,
\end{eqnarray}
where $G_{E,M}^{(I=1)}$ is the isovector form factor.
At forward angles, the asymmetry is sensitive to a combination of
strange electric and magnetic form factors shown in Fig.~11 for
$Q^2=0.48$~GeV$^2$.
With a soft $KNY$ form factor the contributions to both $G_E^s$ and
$G_M^s$ are small and slightly positive~\cite{MM}, in agreement with
the trend of the data.
This result is consistent with the earlier experiment by the
SAMPLE Collaboration at MIT-Bates~\cite{SAMPLE},
$G_M^s = + 0.23 \pm 0.44$ at $Q^2 = 0.1$ GeV$^2$ in a similar
experiment but at backward angles.

Although the experimental results on non-perturbative strangeness
in both structure functions and form factors are still consistent
with zero, they are nevertheless compatible with a soft kaon cloud
around the nucleon.
Future data on $G_{E/M}^s$ from the HAPPEX-II and G0 experiments at
Jefferson Lab with smaller error bars and over a large range of $Q^2$,
as well as the remaining data on the proton and deuteron from SAMPLE,
will hopefully provide conclusive evidence for the presence or otherwise
of a tangible non-perturbative strange component in the nucleon.

\subsection{Polarized Quarks}

Most of the discussion thus far has dealt with the flavor dependence
of quarks in the nucleon.
On the other hand, there has been considerable interest over the past
decade in how the spin of the nucleon is distributed amongst its
constituents~$^{102-104}$.
Spin degrees of freedom allow access to information about the structure
and interactions of hadrons which would otherwise not be available
through unpolarized processes.
Indeed, experiments involving spin-polarized beams and targets have
often yielded surprising results and presented severe challenges to
existing theories.

A fundamental sum rule for the spin in the nucleon states that:
\begin{eqnarray}
{ 1 \over 2 } &=& J_q\ +\ J_g\ ,
\end{eqnarray}
where $J_q$ and $J_g$ are the total quark and gluon angular momenta,
which can be decomposed into their helicity and orbital contributions:
\begin{eqnarray}
\label{Jq}
J_q &=& {1 \over 2} \Delta\Sigma\ +\ L_q\ ,	\\
\label{Jg}
J_g &=& \Delta G\ + L_g\ .
\end{eqnarray}
In particular, $\Delta\Sigma$, which is defined as the forward matrix
element of the axial current,
$\Delta\Sigma = \langle N | \bar\psi \gamma_3 \gamma_5 \psi | N \rangle$,
measures the total helicity of the nucleon carried by quarks, which for
three flavors is:
\begin{eqnarray}
\Delta\Sigma &=& \Delta u + \Delta d + \Delta s\ .
\end{eqnarray}
In non-relativistic quark models the spin of the nucleon is carried
entirely by valence quarks, so that $\Delta\Sigma_{\rm NRQM} = 1$.

The gluon helicity, $\Delta G$, can be measured in high-energy
polarized proton--proton collisions, via charm production through
quark--gluon fusion, or in production of jets with high transverse
momentum~\cite{MALLOT}.
The orbital contributions, $L_q$ and $L_g$, can in principle be extracted
from measurements of off-forward parton distributions in DVCS, or
deeply-virtual meson production experiments~\cite{OFPD}.
Currently there is little empirical information on the gluon polarization,
and the angular momentum distributions are totally unknown.
Note that each term in Eqs.(\ref{Jq}) and (\ref{Jg}) is renormalization
scheme and scale dependent, and only the quark helicity can be defined in
a gauge-invariant manner~\cite{OFPD}.

Experimentally, $\Delta\Sigma$ (which is also referred to as the singlet
axial charge) can be determined from a combination of triplet and octet
axial charges,
\begin{eqnarray}
g_3 &=& \Delta u - \Delta d\ =\ g_A\ ,		\\
g_8 &=& \Delta u + \Delta d\ - \Delta s\ ,
\end{eqnarray}
which are determined from $\beta$-decays of the nucleon and hyperons,
and the spin-dependent $g_1$ structure function of the nucleon.
In the $\overline{\rm MS}$ scheme, the lowest moment of the $g_1$
structure function (at lowest order in perturbative QCD) is
given by~\cite{WEIGL,LR,JG1}:
\begin{eqnarray}
\label{mu2}
\int_0^1 dx g_1^{p(n)}(x,Q^2)
&=& \left[ 1 - \left( {\alpha_s \over \pi} \right)	
    \right]   
\left( \pm {1 \over 12} g_3\
	+\ {1 \over 36} g_8\
	+\ {1 \over 9} \Delta\Sigma
\right),
\end{eqnarray}
where the $\pm$ refers to the proton or neutron.

The first spin structure function experiments at CERN~\cite{EMCSPIN}
suggested a rather small value for $\Delta\Sigma$, in fact consistent
with zero, which prompted the so-called `proton spin-crisis'.
A decade of subsequent measurements of inclusive spin structure
functions using proton, deuteron and $^3$He targets have~\cite{A1}
determined $\Delta\Sigma$ much more accurately, with the current
world average value~\cite{LR} (at a scale of $Q^2=10$~GeV$^2$ in the
$\overline{\rm MS}$ scheme) being $\Delta\Sigma \approx 0.3$.

While the spin fractions carried by quarks in the nucleon require only
the first moment of the inclusive spin-dependent structure functions,
to determine the $x$ dependence of the polarized distributions requires
independent linear combinations of $\Delta q$ which at present can only
be obtained from semi-inclusive scattering.
Generalizing Eq.(\ref{sidis}) to the production of hadrons with a
polarized beam and target, $\vec e \vec N \rightarrow e' h X$, the
difference between the number of hadrons produced at a given $x$, $z$
and $Q^2$ with electron and nucleon spins parallel and antiparallel is
(at leading order) given by:
\begin{eqnarray}
\Delta N^h\ \equiv\
N^h_{\uparrow\Uparrow-\downarrow\Uparrow}(x,z,Q^2)
&\sim& \sum_q e_q^2\ \Delta q(x,Q^2)\ \Delta D_q^h(z,Q^2)\ ,
\end{eqnarray}
where the polarized fragmentation function $\Delta D_q^h$ gives the
probability for a polarized quark $q$ to hadronize into a hadron $h$.

In analogy with the large-$x$ behavior of the unpolarized $u$ and $d$
distributions in Section~4.1, the $x \rightarrow 1$ limit of polarized
quark distributions provides a sensitive test of various mechanisms of
spin-flavor symmetry breaking.
For SU(6) symmetry, the ratio of the polarized to unpolarized quark
distributions is:
\begin{eqnarray}
\label{pol_su6}
{ \Delta u \over u } &=& { 2 \over 3 }\ ,\ \ \ \ \
{ \Delta d \over d }\ =\ - { 1 \over 3 }\ \ \ \ \
[{\rm SU(6)\ symmetry}]\ .
\end{eqnarray}
If the symmetry is broken through the suppression of the $S=1$ diquark
contributions in the nucleon, then in the limit $x \rightarrow 1$:
\begin{eqnarray}
\label{pol_s0}
{ \Delta u \over u } &\longrightarrow& 1\ ,\ \ \ \ \
{ \Delta d \over d }\ \longrightarrow\ - { 1 \over 3 }\ \ \ \
[S=0\ {\rm dominance}]\ .
\end{eqnarray}
The perturbative QCD prediction (where the dominant configurations of the
proton wave function are those in which the spins of the interacting quark
and proton are aligned) on the other hand, is:
\begin{eqnarray}
\label{pol_pqcd}
{ \Delta u \over u } &\longrightarrow& 1\ ,\ \ \ \ \
{ \Delta d \over d }\ \longrightarrow\ 1\ \ \ \ \
[S_z=0\ {\rm dominance}]\ .
\end{eqnarray}
Note that the predictions for the $\Delta d$ quark in particular are
quite different in the perturbative and non-perturbative models, even
differing by sign.

The spin-flavor distributions can be directly measured via polarization
asymmetries for the difference between $\pi^+$ and $\pi^-$ production
cross sections on the proton~\cite{SEMIDIS,HERMES_SEMISPIN}:
\begin{eqnarray}
A_p^{\pi^+ - \pi^-}
&=& { \Delta N_p^{\pi^+} - \Delta N_p^{\pi^-} \over
      \Delta N_p^{\pi^+} + \Delta N_p^{\pi^-} }\
 =\ { 4 \Delta u_{\rm val} - \Delta d_{\rm val} \over
      4 u_{\rm val} - d_{\rm val} }\ ,
\end{eqnarray}
where the dependence on fragmentation functions and sea quarks cancels.
A combination of inclusive and semi-inclusive asymmetries using protons
and deuterons at an energy upgraded Jefferson Lab will allow the
spin-dependent $\Delta u$ and $\Delta d$ distributions to be determined
up to $x \sim 0.8$ with good statistics~\cite{FOREST}, which should be
able to discriminate between the various model scenarios in
Eqs.(\ref{pol_su6})-(\ref{pol_pqcd}).

At smaller $x$, similar combinations of asymmetries could also be able
to measure the polarized antiquark distributions, $\Delta \bar d$
and $\Delta \bar u$.
These are particularly interesting in view of the qualitatively
different predictions in non-perturbative models.
While chiral (pion) cloud models do not allow any polarization in the
antiquark sea, the Pauli exclusion principle on the other hand predicts
quite a large asymmetry,
$(\Delta \bar u - \Delta \bar d)/(\bar d - \bar u) = 5/3$,
even bigger than in the unpolarized sea~\cite{SST,POLSEA}.
Measurement of the polarized asymmetry in semi-inclusive scattering
would then enable the relative sizes of the pion and Pauli blocking
contributions to $\bar d-\bar u$ to be disentangled.

While data on the polarized quark distributions has slowly been
accumulating from various experiments, and plans are under way to
systematically measure the polarization of the gluons, until recently
there has been very little discussion about the fraction of the nucleon
spin residing in angular momentum~\cite{ORB}.
This changed somewhat when it was demonstrated~\cite{OFPD} that the
orbital angular momentum contributions could be determined from
off-forward parton distributions measured in deeply-virtual Compton
scattering (see Section~3.4).

In particular, it was shown~\cite{OFPD} that a sum rule can be derived
relating moments of the OFPDs to the total angular momentum carried by
quarks and gluons:
\begin{eqnarray}
\label{sum2}
\int_{-1}^1 dx\ x \Big( H(x,\xi,t) + E(x,\xi,t) \Big)
&=& A(t) + B(t)\ , 
\end{eqnarray}
where $A$ and $B$ are form factors of the energy-momentum tensor in
QCD~\cite{JM}:
\begin{eqnarray}
\label{Tmunu_q}
T^{\mu\nu}_q
&=& {1 \over 2}
\left( \overline \psi \gamma^{\{\mu} i \overrightarrow{ D^{\nu\}}} \psi
     + \overline \psi \gamma^{\{\mu} i \overleftarrow{ D^{\nu\}}}  \psi  
\right) \ ,		\\
T^{\mu\nu}_g
&=& {1\over 4}g^{\mu\nu} F^2 - F^{\mu\alpha}F^\nu_{~\alpha} \ .
\end{eqnarray}
for quarks and gluons, respectively, where the braces $\{ \cdots \}$
represent symmetrization of indices.
The matrix elements of $T^{\mu\nu}$ can be expanded as~\cite{OFPD}:
\begin{eqnarray}
\langle P' | T^{\mu\nu} | P \rangle
&=& \overline u(P')
\Big[ {1 \over 2}  A(t)\ \gamma^{\{\mu} (P+P')^{\nu\}}\	   \nonumber\\
& & \hspace*{-1cm}
 +\ {1 \over 4M} B(t)\ (P+P')^{\{\mu}
	i\sigma^{\nu\}\alpha} (P'-P)_\alpha
   + \cdots
\Big] u(P)\ .
\end{eqnarray}  
One can then show that the total angular momentum carried by
quarks is given by:
\begin{eqnarray}
J_q &=& {1\over 2} \Big( A(0) + B(0) \Big)\ .
\end{eqnarray}
Combining the extracted $J_q$ with $\Delta\Sigma$ measured in inclusive
DIS, one can then determine the orbital angular momentum of the quarks
in the nucleon.
An analogous sum rule can also be written for the total gluon angular
momentum, $J_g$, which can be obtained from OFPDs measured in
deeply-virtual meson production.
{}From this the gluon orbital angular momentum can be extracted once
the gluon helicity $\Delta G$ is known.

The program to measure the off-forward parton distributions
$H, E, \cdots$ in deeply-virtual Compton scattering and meson
production experiments is difficult, requiring a large coverage
of kinematics and knowledge of background such as the Bethe-Heitler
process~\cite{DVCS} for DVCS.
The first steps along the road to mapping out these fundamental
quantities are already being taken at Jefferson Lab and HERMES.

\section{Conclusion}

Thanks to recent advances in accelerator technology that have enabled 
precise data to be collected at the world's particle accelerators, we have
been able to probe the fascinating inner structure of the nucleon with
unprecedented clarity.
Though much has been learned from inclusive DIS experiments, future
analyses of nucleon structure will focus more on semi-inclusive reactions,
which will enable the spin and flavor composition of protons and neutrons
to be resolved with greater precision.
Furthermore, there is a growing appreciation of the need to understand
the common underlying physics revealed through a range of observables,
from elastic form factors to deep-inelastic structure functions.

Some of the most exciting recent developments have been in the study of
the non-perturbative structure of the proton sea through asymmetries in
sea quark distributions, which illustrate the relevance of chiral symmetry
breaking in QCD even at high energies.
Important breakthroughs in our understanding of the proton spin have
opened the way to accessing for the first time information about the
full helicity and orbital momentum distributions in the nucleon.
In addition, perhaps longest overdue is the need to determine the valence
quark distributions in the region of large Bjorken-$x$, which should
settle the long-standing puzzle of the precise $x \rightarrow 1$ behavior
of structure functions and shed light on the mechanisms of spin-flavor
symmetry breaking in the nucleon.

To make the inroads necessary to achieve a deeper understanding of these
issues will require full utilization of the high luminosities and machine
duty factors at modern accelerator facilities such as Jefferson Lab.
We can anticipate the new generations of experiments to reveal much more
of the intriguing world of subnucleon dynamics.

\newpage
\section*{Acknowledgments}

I would like to thank Jose Goity for organizing an excellent HUGS
summer school.
This work was supported by the Australian Research Council and
the U.S. Department of Energy contract \mbox{DE-AC05-84ER40150}.
under which the Southeastern Universities Research Association (SURA)
operates the Thomas Jefferson National Accelerator Facility
(Jefferson Lab).



\section*{References}

\begin{thebibliography}{99}

\bibitem{YUKAWA}
H.~Yukawa,
{\em Proc. Phys. Math. Soc. Japan} {\bf 17}, 48 (1935).

\bibitem{WICK}
G.C.~Wick,
{\em Nature} {\bf 142}, 993 (1938).

\bibitem{QUARKS}
M.~Gell-Mann,
\Journal{\PL}{8}{214}{1964};
%
G.~Zweig,
CERN preprint 8419/TH 412 (1964).

\bibitem{COLOR}
O.W.~Greenberg,
\Journal{\PRL}{13}{598}{1964};
%
M.Y.~Han and Y.~Nambu,
\Journal{\PR}{139}{1006}{1965}.

\bibitem{EARLYSLAC}
W.K.H.~Panofsky,
in {\em Proceedings of the 14th International Conference on High Energy
Physics}, Vienna, 1968 (CERN Scientific Information Service, Geneva);
%
E.D.~Bloom et al.,
\Journal{\PRL}{23}{930}{1969};
%
J.I.~Friedman and H.W.~Kendall,
\Journal{\ARNPS}{22}{203}{1972}.

\bibitem{QCD}
H.~Fritzsch and M.~Gell-Mann,
in {\em Proceedings of the 16th. International
Conference on High Energy Physics},
ed. J.P.~Jackson and A.~Roberts
(National Accelerator Lab., Batavia, Illinois, 1972);
%
H.~Fritzsch, M.~Gell-Mann and H.~Leutwyler,
\Journal{\PLB}{47}{365}{1973};
%
D.J.~Gross and F.~Wilczek,
\Journal{\PRD}{8}{3633}{1973}.

\bibitem{MOLECULE}
N.~Isgur,
in {\em Proceedings of the 26th International Conference on
High Energy Physics (ICHEP 92)}, Dallas, Texas, August 1992.

\bibitem{BURAS}
A.J.~Buras,
\Journal{\RMP}{52}{199}{1980}.

\bibitem{YNDURAIN}
F.J.~Yndurain,
{\em Quantum Chromodynamics}
(Springer-Verlag, Berlin, 1983).

\bibitem{MUTA}
T.~Muta,
{\em Foundations of Quantum Chromodynamics}
(World Scientific, Singapore, 1987). 

\bibitem{HANDBOOK}
G.~Sterman et al.,
\Journal{\RMP}{67}{157}{1995}.

\bibitem{LATTICE}
K.G.~Wilson,
\Journal{\PRD}{10}{2445}{1974};
%
D.G.~Richards,
these proceedings,
nucl-th/0006020.

\bibitem{CHIPT}
S.~Weinberg,
{\em Physica (Amsterdam)} {\bf 96 A}, 327 (1979);
%
J.~Gasser and H.~Leutwyler,   
\Journal{\ANN}{158}{142}{1984};
%
B.R.~Holstein,
hep-ph/9911449,
and these proceedings.

\bibitem{HEY}
A.J.G.~Hey and J.E.~Mandula,
\Journal{\PRD}{5}{2610}{1972}.

\bibitem{LP}
E.~Leader and E.~Predazzi,
{\em An Introduction to Gauge Theories and Modern Particle Physics}
(Cambridge Univ. Press, 1996).

\bibitem{IOANA}
I.~Niculescu,
private communication.

\bibitem{FFP}
R.~Hofstadter and R.W.~McAllister,
\Journal{\PR}{98}{183}{1955};
%
R.W.~Mcallister and R.~Hofstadter,
\Journal{\PR}{102}{851}{1956}.
%
R.~Hofstadter et al.,
\Journal{\PRL}{5}{263}{1960}.

\bibitem{LB}
G.P.~Lepage and S.J.~Brodsky,
\Journal{\PRL}{43}{545}{1979};
\Journal{\PRD}{22}{2157}{1980}.

\bibitem{FEYN69}
R.P.~Feynman,
in {\em Proceedings of the 3rd Topical Conference on High Energy
Collisions of Hadrons}, Stony Brook, NY, ed. C.N.~Yang et al.
(Gordon \& Breach, New York, 1969);
%
\Journal{\PRL}{23}{1415}{1969}.

\bibitem{BP}
J.D.~Bjorken and E.A.~Paschos,
\Journal{\PR}{185}{1975}{1969}.

\bibitem{OPE}
K.~Wilson,
\Journal{\PR}{179}{1499}{1969};
%
R.A.~Brandt and G.~Preparata, 
\Journal{\NPB}{27}{541}{1971};
%
O.~Nachtmann, 
\Journal{\NPB}{63}{237}{1973}.

\bibitem{JAF83}
R.L.~Jaffe,   
\Journal{\NPB}{229}{205}{1983}.

\bibitem{WEIGL}
T.~Weigl and W.~Melnitchouk,
\Journal{\NPB}{465}{267}{1996}.

\bibitem{MODELS}
A.~Le~Yaouanc, L.~Oliver, O.~P\`ene and J.C.~Reynard,
\Journal{\PRD}{9}{2636}{1974};
%
R.L.~Jaffe,
\Journal{\PRD}{11}{1953}{1975};
%
N.~Cabbibo and R.~Petronzio,
\Journal{\NPB}{137}{395}{1978};
%
C.J.~Benesh and G.A.~Miller,
\Journal{\PRD}{36}{1344}{1987}.

\bibitem{BAG}
A.I.~Signal and A.W.~Thomas,
\Journal{\PRD}{40}{2832}{1989};
%
A.W.~Schreiber, A.W.~Thomas and J.T.~Londergan,
\Journal{\PLB}{237}{120}{1989}.

\bibitem{SST}
A.W.~Schreiber, A.I.~Signal and A.W.~Thomas,
\Journal{\PRD}{44}{2653}{1991}.

\bibitem{JAFROS}
R.L.~Jaffe and G.G.~Ross,
\Journal{\PL}{93 B}{313}{1980}.

\bibitem{PARPETR}
G.~Parisi and R.~Petronzio
\Journal{\PL}{62 B}{331}{1976}.

\bibitem{GP74}
H.~Georgi and H.D.~Politzer,
\Journal{\PRD}{9}{416}{1974}.

\bibitem{GW74}
D.J.~Gross and F.~Wilczek,
\Journal{\PRD}{9}{920}{1974}.

\bibitem{DGLAP}
Yu.L.~Dokshitzer,
\Journal{\JETP}{46}{641}{1977};
%
V.N.~Gribov and L.N.~Lipatov,
\Journal{\SJNP}{15}{439}{1972};
%
L.N.~Lipatov,
\Journal{\SJNP}{20}{181}{1974};
%
G.~Altarelli and G.~Parisi,
\Journal{\NPB}{126}{278}{1977}.

\bibitem{CTEQ}
H.L.~Lai et al.,
\Journal{\EPJC}{12}{375}{2000}.

\bibitem{MRST}
A.D.~Martin, R.G.~Roberts, W.J.~Stirling and R.S.~Thorne,
\Journal{\EPJC}{4}{463}{1998}.

\bibitem{GRV}
M.~Gluck, E.~Reya and A.~Vogt,
\Journal{\EPJC}{5}{461}{1998}.

\bibitem{SEMI}
P.V.~Landshoff and J.C.~Polkinghorne,
\Journal{\NPB}{33}{221}{1971};
%
J.~Ellis,
\Journal{\PLB}{35}{537}{1971};
%
J.D.~Stack,
\Journal{\PRL}{28}{57}{1972};
%
T.~Sloan, G.~Smadja and R.~Voss,
\Journal{\PRP}{162}{45}{1980};
%
R.~Renton and W.S.C.~Williams,
\Journal{\ARNPS}{31}{193}{1981}.

\bibitem{EPIC}
W.~Melnitchouk,
in {\em Proceedings of the Workshop on Physics with a High-Luminosity
Polarized Electron-Ion Collider}, IUCF, April 1999,
hep-ph/9906488.

\bibitem{HERMES_DU}
K.~Ackerstaff,
First Results from the HERMES Experiment using Unpolarized Targets,
PhD thesis, Univ. Hamburg, 1996.

\bibitem{EMCFRAG}
J.J.~Aubert et al.,
\Journal{\PLB}{110}{73}{1982};
\Journal{\NPB}{213}{213}{1983}.

\bibitem{OFPD}
X.~Ji,
\Journal{\PRL}{78}{610}{1997};
\Journal{\JPG}{24}{1181}{1998}.

\bibitem{NFPD}
A.V.~Radyushkin,
\Journal{\PRD}{56}{5524}{1997}.

\bibitem{JJ}
R.L.~Jaffe and X.~Ji,
\Journal{\PRL}{67}{552}{1991};
%
A.V.~Manohar,
\Journal{\PRL}{66}{289}{1991}.

\bibitem{OFPDBAG}
X.~Ji, W.~Melnitchouk and X.~Song,
\Journal{\PRD}{56}{5511}{1997}.

\bibitem{CTEQ_LX}
S.~Kuhlmann et al.,
\Journal{\PLB}{476}{291}{2000}.

\bibitem{MTNP}
W.~Melnitchouk and A.W.~Thomas,   
\Journal{\PLB}{377}{11}{1996}.

\bibitem{CLOSE79}
F.E.~Close,
{\em An Introduction to Quarks and Partons}
(Academic Press, 1979).

\bibitem{FEYN72}
R.P.~Feynman,
{\em Photon Hadron Interactions}
(Benjamin, Reading, Massachusetts, 1972).

\bibitem{CLOSE73}
F.E.~Close,
\Journal{\PLB}{43}{422}{1973}.

\bibitem{CARLITZ}
R.~Carlitz,
\Journal{\PLB}{58}{345}{1975}.

\bibitem{CT}
F.E.~Close and A.W.~Thomas,
\Journal{\PLB}{212}{227}{1988}.

\bibitem{ISGUR_V}
N.~Isgur,
\Journal{\PRD}{59}{034013}{1999}.

\bibitem{FJ}
G.R.~Farrar and D.R.~Jackson,
\Journal{\PRL}{35}{1416}{1975}.

\bibitem{SMEAR}
W.~Melnitchouk, A.W.~Schreiber and A.W.~Thomas,
\Journal{\PLB}{335}{11}{1994};
\Journal{\PRD}{49}{1183}{1994}.

\bibitem{W}
W.~Melnitchouk and J.C.~Peng,
\Journal{\PLB}{400}{220}{1997}.

\bibitem{SEMID}
W.~Melnitchouk, M.~Sargsian and M.I.~Strikman,
\Journal{\ZPA}{359}{99}{1997}.

\bibitem{PARITY}
P.~Souder,
in {\em Proceedings of Workshop on CEBAF at Higher Energies},
CEBAF, Newport News, 1994;
%
R.~Michaels,
in {\em Physics and Instrumentation with 6-12 GeV Beams},
Jefferson Lab, Newport News, June 1998.

\bibitem{HERACC}
H1 Collaboration, C.~Adloff et al.,
DESY-99-107, hep-ex/9908059;
%
ZEUS Collaboration, J.~Breitweg et al.,
DESY-99-059, hep-ex/9907010.

\bibitem{H3}
G.G.~Petratos et al.,
in {\em Proceedings of Workshop on Experiments with Tritium at JLab},
Jefferson Lab, Newport News, Virginia, September 1999;
%
I.R.~Afnan et al.,
nucl-th/0006003.

\bibitem{SEMIPI}
W.~Melnitchouk, J.~Speth and A.W.~Thomas,
\Journal{\PLB}{435}{420}{1998}.

\bibitem{BG}
E.D.~Bloom and F.J.~Gilman,
\Journal{\PRL}{16}{1140}{1970}.

\bibitem{F2JL}
I.~Niculescu,  
Ph.D. thesis, Hampton University, 1999;
C.~Keppel,
talk presented at 7th International Workshop on Deep Inelastic
Scattering and QCD (DIS 99),
Zeuthen, Germany, Apr. 1999.

\bibitem{DY}
S.D.~Drell and T.-M.~Yan,
\Journal{\PRL}{24}{181}{1970}.

\bibitem{WEST}
G.B.~West,
\Journal{\PRL}{24}{1206}{1970}.

\bibitem{MEL}
W.~Melnitchouk,
\Journal{\PRL}{86}{35}{2001}.

\bibitem{RUJ}
A.~De R\'ujula, H.~Georgi and H.D.~Politzer,
\Journal{\ANN}{103}{315}{1975}.   

\bibitem{IJMV}
N.~Isgur, S.~Jeschonnek, W.~Melnitchouk and J.W. Van~Orden,
\Journal{\PRD}{64}{054005}{2001}.

\bibitem{CM}
C.E.~Carlson and N.C.~Mukhopadhyay,
\Journal{\PRD}{58}{094029}{1998};
\Journal{\PRD}{41}{2343}{1989}.

\bibitem{MDM}
P.~Mergell, U.-G.~Mei\ss ner and D.~Drechsel,
\Journal{\NPA}{596}{367}{1996}.

\bibitem{GEMJL}
M.K.~Jones et al.,
\Journal{\PRL}{84}{1398}{2000}.

\bibitem{SCIENCE}
A.~Watson,
{\em Science} {\bf 283}, 472 (1999).

\bibitem{NMCGSR}
P.~Amaudraz et al.,
\Journal{\PRL}{66}{2712}{1991}.

\bibitem{NA51}
A.~Baldit et al.,
\Journal{\PLB}{332}{244}{1994}.

\bibitem{E866}
E.A.~Hawker et al.,
\Journal{\PRL}{80}{3715}{1998}.

\bibitem{MTSHAD}
W.~Melnitchouk and A.W.~Thomas,
\Journal{\PRD}{47}{3783}{1993}.

\bibitem{TM}
A.W.~Thomas and W.~Melnitchouk,
\Journal{\NPA}{631}{296}{1998}.

\bibitem{AWT83}
A.W.~Thomas, 
\Journal{\PLB}{126}{97}{1983}.

\bibitem{CBM}
S.~Th\'eberge, G.A.~Miller and A.W.~Thomas,
\Journal{\PRD}{22}{2838}{1980};
%
A.W.~Thomas,
\Journal{\ANP}{13}{1}{1984};
%
D.H.~Lu, A.W.~Thomas and A.G.~Williams,
\Journal{\PRC}{57}{2628}{1998}.

\bibitem{GI}
P.~Geiger and N.~Isgur,
\Journal{\PRD}{55}{299}{1997}.

\bibitem{IMF}  
S.~Weinberg,
\Journal{\PR}{150}{1313}{1966};
%
S.D.~Drell, D.J.~Levy, and T.M.~Yan,
\Journal{\PRD}{1}{1035}{1970}.

\bibitem{REV}
J.~Speth and A.W.~Thomas,
\Journal{\ANP}{24}{83}{1998}.

\bibitem{SULL}
J.D.~Sullivan,
\Journal{\PRD}{5}{1732}{1972}.

\bibitem{ZOL}
V.R.~Zoller,
\Journal{\ZPC}{54}{425}{1992};
\Journal{\ZPC}{60}{141}{1993}.

\bibitem{MTV}
W.~Melnitchouk and A.W.~Thomas,
\Journal{\PRD}{47}{3794}{1993}.  

\bibitem{HEAVY}
W.~Melnitchouk and A.W.~Thomas,
\Journal{\PRD}{47}{3794}{1993}; 
%
H.~Holtmann, A.~Szczurek and J.~Speth,
\Journal{\NPA}{596}{631}{1996}.

\bibitem{AXIAL}
T.~Kitagaki et al.,
\Journal{\PRD}{42}{1331}{1990}.

\bibitem{DYN}   
W.~Melnitchouk, J.~Speth and A.W.~Thomas,
\Journal{\PRD}{59}{014033}{1999}.

\bibitem{FMST}
F.M.~Steffens and A.W.~Thomas,
\Journal{\PRC}{55}{900}{1997}.

\bibitem{FF}
R.D.~Field and R.P.~Feynman,
\Journal{\PRD}{15}{2590}{1977}.

\bibitem{LMS}
J.~Levelt, P.J.~Mulders and A.W.~Schreiber,
\Journal{\PLB}{263}{498}{1991}.

\bibitem{HERMES_DUBAR}
K.~Ackerstaff et al.,
\Journal{\PRL}{81}{5519}{1998}.

\bibitem{EXCL}
W.~Melnitchouk,
in {\em Proceedings of the Workshop on Exclusive \& Semi-Exclusive
Processes at High Momentum Transfer}, Jefferson Lab, June 1999,
hep-ph/9909463.

\bibitem{FFN}
S.~Platchkov et al.,
\Journal{\NPA}{510}{740}{1990};
%
P.~Stoler,
\Journal{\PRP}{226}{103}{1993}.

\bibitem{IKK}
N.~Isgur, G.~Karl and R.~Koniuk,
\Journal{\PRL}{41}{1269}{1978}.

\bibitem{IKS}
N.~Isgur, G.~Karl and D.W.L.~Sprung,
\Journal{\PRD}{23}{163}{1981}.

\bibitem{ISGUR_G}
N.~Isgur,
\Journal{\PRL}{83}{272}{1999}.

\bibitem{JT}
X.~Ji and J.~Tang,
\Journal{\PLB}{362}{182}{1995}.

\bibitem{CCFR}
A.O.~Bazarko et al.,
\Journal{\ZPC}{65}{189}{1995}.

\bibitem{ST}
A.I.~Signal and A.W.~Thomas,
\Journal{\PLB}{191}{206}{1987}.

\bibitem{MM}
W.~Melnitchouk and M.~Malheiro,
\Journal{\PRC}{55}{431}{1997};
\Journal{\PLB}{451}{224}{1999};
\Journal{\PRC}{56}{2373}{1997}.

\bibitem{LNA}
A.W.~Thomas, W.~Melnitchouk and F.M.~Steffens,
\Journal{\PRL}{85}{2892}{2000}.

\bibitem{SAMPLE}
B.~Mueller et al.,
\Journal{\PRL}{78}{3824}{1997}.

\bibitem{HAPPEX}
K.A.~Aniol et al.,
\Journal{\PRL}{82}{1096}{1999}.

\bibitem{CR}
F.E.~Close and R.G.~Roberts,
\Journal{\PLB}{316}{165}{1993}.

\bibitem{BASS}
S.D.~Bass and A.W.~Thomas,
\Journal{\PPNP}{33}{449}{1994}.
%
S.D.~Bass,
\Journal{\EPJA}{5}{17}{1999}.

\bibitem{LR}
B.~Lampe and E.~Reya,
hep-ph/9810270.

\bibitem{MALLOT}
G.K.~Mallot,
\Journal{\JPG}{25}{1539}{1999}.

\bibitem{JG1}
X.~Ji and P.~Unrau,
\Journal{\PLB}{333}{228}{1994};
%
X.~Ji and W.~Melnitchouk,
\Journal{\PRD}{56}{1}{1997}. 

\bibitem{EMCSPIN}
J.~Ashman et al.,
\Journal{\PLB}{206}{364}{1988}.

\bibitem{A1}
K.~Abe et al.,
\Journal{\PRL}{79}{26}{1997};
%
B.~Adeva et al.,
\Journal{\PRD}{58}{112001}{1998};
%
K.~Abe et al.,
{\em ibid} D {\bf 58} 112003 (1998).

\bibitem{SEMIDIS}
L.L.~Frankfurt, M.I.~Strikman, L.~Mankiewicz, A.~Schafer, E~Rondio,
A.~Sandacz and V.~Papavassiliou,
\Journal{\PLB}{230}{141}{1989}.

\bibitem{HERMES_SEMISPIN}
K.~Ackerstaff,
\Journal{\PLB}{464}{123}{1999}.

\bibitem{FOREST}
T.A.~Forest,
private communication.

\bibitem{POLSEA}
B.~Dressler, K.~Goeke, M.V.~Polyakov and C.~Weiss,
\Journal{\EPJC}{14}{147}{2000};
%
M.~Gl\"uck and E.~Reya,
hep-ph/0002182;
%
R.S.~Bhalerao,
hep-ph/0003075.

\bibitem{ORB}
L.~Sehgal,
\Journal{\PRD}{10}{1663}{1974};
%
X.~Ji, J.~Tang and P.~Hoodbhoy,
\Journal{\PRL}{76}{740}{1996};
%
S.V.~Bashinsky and R.L.~Jaffe,
\Journal{\NPB}{536}{303}{1998};
%
N.~Mathur, S.J.~Dong, K.F.~Liu, L.~Mankiewicz and N.C.~Mukhopadhyay,
hep-ph/9912289.

\bibitem{JM}   
R.L.~Jaffe and A.~Manohar,
\Journal{\NPB}{337}{509}{1990}.

\bibitem{DVCS}
M.~Vanderhaeghen, P.A.M.~Guichon and M.~Guidal,
\Journal{\PRL}{80}{5064}{1998};
\Journal{\PRD}{60}{094017}{1999}.

\end{thebibliography}
\end{document}